\documentclass[sigplan,screen]{acmart}

\AtBeginDocument{%
  }

\setcopyright{cc}
\setcctype{by}
\acmDOI{10.1145/3696443.3708932}
\acmYear{2025}
\copyrightyear{2025}
\acmISBN{979-8-4007-1275-3/25/03}
\acmConference[CGO '25]{Proceedings of the 23rd ACM/IEEE International Symposium on Code Generation and Optimization}{March 01--05, 2025}{Las Vegas, NV, USA}
\acmBooktitle{Proceedings of the 23rd ACM/IEEE International Symposium on Code Generation and Optimization (CGO '25), March 01--05, 2025, Las Vegas, NV, USA}
\received{2024-09-09}
\received[accepted]{2024-11-04}

\usepackage{microtype}

\usepackage{subcaption}
\usepackage{graphicx}

\usepackage{wrapfig}
\usepackage[inline]{enumitem}
\usepackage{multicol}
\usepackage{semantic}

\usepackage{balance}

\usepackage{simplebnf}
\usepackage{expl3}
\usepackage{xparse}
\usepackage{mathtools}
\usepackage{tabularray}
\usepackage{multirow}

\usepackage{color,colortbl}
\usepackage{xcolor}

\definecolor{myyellow}{rgb}{1, 1, 0}
\definecolor{mygray}{rgb}{0.851, 0.851, 0.851}

\usepackage{paracol}

\usepackage{algorithm}
\usepackage[noend]{algpseudocode}
\algtext*{EndIf}

\algdef{SE}[DOWHILE]{Do}{doWhile}{\algorithmicdo}[1]{\algorithmicwhile\ #1}%

\algnewcommand\algorithmicsmatch{\textbf{match}}
\algnewcommand\algorithmicsis{\textbf{is}}
\algnewcommand\algorithmiccase{\textbf{case}}
\algnewcommand\algorithmicassert{\texttt{assert}}
\algnewcommand\algorithmicpass{\textbf{pass}}
\algnewcommand\Pass{\algorithmicpass}
\algnewcommand\Assert[1]{\State \algorithmicassert(#1)}%
\algdef{SE}[MATCH]{Match}{EndMatch}[1]{\algorithmicsmatch\ #1\ \algorithmicsis}{\algorithmicend\ \algorithmicsmatch}%
\algdef{SE}[CASE]{Case}{EndCase}[1]{\algorithmiccase\ \text{'}#1\text{'}:}{\algorithmicend\ \algorithmiccase}%
\algtext*{EndMatch}%
\algtext*{EndCase}%

\algrenewcommand\algorithmicindent{0.8em}%

\usepackage{tikz}
\usetikzlibrary{shapes.geometric, shapes.symbols, shapes.misc, arrows, positioning, backgrounds, calc, decorations.pathmorphing,arrows.meta}

\usepackage{listings}
\definecolor{codegreen}{rgb}{0,0.6,0}
\definecolor{codegray}{rgb}{0.5,0.5,0.5}
\definecolor{codepurple}{rgb}{0.58,0,0.82}
\definecolor{backcolour}{rgb}{0.95,0.95,0.92}
\definecolor{myblue}{rgb}{0.0, 0.0, 1.0}

\ccsdesc[500]{Software and its engineering~Compilers}
\ccsdesc[500]{Software and its engineering~Automated static analysis}

\keywords{points-to analysis, static analysis, pointer analysis, compiler, optimization} 

\begin{document}

\title{SkipFlow: Improving the Precision of Points-to Analysis using Primitive Values and Predicate Edges}

\newcommand\predEnabled[0]{$\mathit{pred}^{\mathit{on}}$}
\newcommand\analysisName[0]{SkipFlow}

\newcommand{\mynote}[3]{
\definecolor{tempcolor}{RGB}{#2}
{\color{tempcolor}\fbox{\bfseries\sffamily\scriptsize#1}#3\fbox{\bfseries\sffamily\scriptsize}}
}

\newcommand\TODO[1]{\mynote{TODO}{255,64,0}{#1}}

\newcommand\copied[1]{\mynote{COPIED}{255,0,0}{#1}}
\newcommand\cwi[1]{\mynote{Christian Wimmer}{64,128,0}{#1}}
\newcommand\cs[1]{\mynote{Codrut Stancu}{0,64,128}{#1}}

\newcommand\dk[1]{#1}

\newcommand\tv[1]{\mynote{Tomas}{64,64,128}{#1}}

\newcommand\NUMBER[1]{#1}

\author{David Kozak}
\orcid{0000-0002-8846-922X}
\email{ikozak@fit.vut.cz}
\affiliation{%
  \institution{Oracle Labs and Brno University of Technology}
  \country{Czechia}
}

\author{Codrut Stancu}
\orcid{0009-0007-3646-3663}
\email{codrut.stancu@oracle.com}
\affiliation{%
  \institution{Oracle Labs}
  \country{Switzerland}
}

\author{Tomáš Vojnar}
\orcid{0000-0002-2746-8792}
\email{vojnar@fi.muni.cz}
\affiliation{%
  \institution{Masaryk University and Brno University of Technology}
  \country{Czechia}
}

\author{Christian Wimmer}
\email{cwimmer@acm.org}
\orcid{0009-0003-3213-9306}
\affiliation{%
  \institution{Work done while being a member of Oracle Labs}
  \country{USA}
}


\begin{abstract}
A typical points-to analysis such as Andersen's or Steensgaard's may lose precision because it ignores the branching structure of the analyzed program.
Moreover, points-to analysis typically focuses on objects only, not considering instructions manipulating primitive values.
We argue that such an approach leads to an unnecessary precision loss, for example, when primitive constants \texttt{true} and \texttt{false} flow out of method calls.
We propose a~novel lightweight points-to analysis called \analysisName{} that interprocedurally tracks the flow of both primitives and objects, and explicitly captures the branching structure of the code using predicate edges.
At the same time, however, SkipFlow is as lightweight and scalable as possible, unlike a traditional flow-sensitive analysis.
We apply \analysisName{} to GraalVM Native Image, a closed-world solution
to building standalone binaries for Java applications. We evaluate the
implementation using a~set of microservice applications as well as well-known
benchmark suites. We show that \analysisName{} reduces the size of the application in terms of reachable
methods by 9\% on average without significantly increasing the analysis time.
\end{abstract}

\maketitle

\section{Introduction}
Points-to analysis has been applied in many areas including compiler optimizations~\cite{LLVM:CGO04}, security analysis~\cite{Arzt14,Livshits05}, bug hunting~\cite{Guyer05,Liu18}, escape analysis~\cite{Lundberg09}, call graph construction~\cite{Karim12,Reif16}, and program understanding~\cite{Li16,Fiutem99}. 
However, a typical points-to analysis such as Andersen's~\cite{Andersen92} or Steensgaard's~\cite{Steensgaard96} may loose a~lot of precision because it ignores the branching structure of the analyzed programs~\cite{Hind99}.

A flow-sensitive analysis can mitigate the issue by computing information per program point~\cite{Zhao18}, but such an analysis is known to have scalability issues~\cite{Li11}. This complicates usage in domains such as optimizing compilers or light-weight bug-hunting tools, where the analysis has to finish in \textit{a~few minutes} even on large programs.

Our key observation is that many branching instructions can be efficiently evaluated during the analysis and used to prune out unreachable successor branches without increasing the overhead of the analysis. 
To do so, we introduce the notion of \textit{predicate edges} connecting the branching conditions with nodes representing instructions contained within the branches. In addition, we extend the domain of values commonly tracked by points-to
analysis with \textit{primitive values}, motivated by the observation that many branching instructions are based on primitive values returned from other methods. 

We propose a novel predicated points-to analysis, named \emph{SkipFlow}, that tracks the flow of both primitives and objects
interprocedurally using a~data structure called a \emph{predicated value
propagation graph} (PVPG). We show how the PVPG can be extracted by a~sequential
pass over a~program and present an algorithm computing the \textit{value
states}, i.e., points-to sets, for all variables and fields.

We perform an extensive set of experiments using a~set of modern microservice
applications as well as the well-known benchmarking suites
Renaissance~\cite{renaissance} and DaCapo~\cite{Blackburn06}.

\begin{figure}
\begin{lstlisting}[language=Java]
class Scene {
    void render(..., Display display) {
        if (display == null) {
            display = new FrameDisplay();
        }
        ...
    }
}

class BucketRenderer {
    void render(Display display) {
        ...
        display.imageBegin();
        ...
    }
}
\end{lstlisting}
\caption{A \texttt{DaCapo Sunflow} motivating example.}
\label{fig:motivating-example-dacapo}
\end{figure}

\dk{When applied on top of a~context-insensitive typed-based points-to analysis~\cite{Wimmer24},} \analysisName{} reduces the size of the applications in terms of reachable
methods by \NUMBER{max 9.2\%, min 3.3\%, avg 6.3\% for microservices, max
17.2\%, min 3.7\%, avg 8.4\% for Renaissance, and max 52.3\%, min
3.5\%, avg 13.3\% for DaCapo} without much negative impact on the
analysis time. In fact, \analysisName{} often reduces the analysis time because fewer
methods have to be processed. 

The implementation presented in this paper is based on GraalVM Native Image~\cite{Wimmer19}, which is written in Java, analyzes Java bytecode, and Java is used for all examples in this paper. 
Nevertheless, our approach is not limited to Java or languages that compile to Java bytecode. 
It can be applied to all managed languages that are amenable to points-to analysis, such as C\# or other languages of the .NET framework.

In summary, this paper contributes to the state of the art the following:
\begin{itemize}

    \item We propose a~novel approach that increases the precision of a points-to analysis using
    \textit{predicate edges}, in which flow-sensitivity is maintained for selected kinds of branching instructions only.

    \item We extend the points-to analysis to also track primitive constant values.

    \item We present an implementation for GraalVM Native Image and evaluate it using a~set of modern microservice applications as well
    as the well-known benchmark suites Renaissance and DaCapo. The results show that
    \analysisName{} reduces the size of the applications in terms of reachable methods
    by \NUMBER{9\%}, and also reduces analysis time in many cases.

\end{itemize}

\section{Real-World Motivating Examples}
\label{sec:motivatingExample}

Consider Figure~\ref{fig:motivating-example-dacapo} that is taken from the
\texttt{DaCapo Sunflow} benchmark~\cite{Blackburn06}.  The method
\texttt{Scene.render} has a parameter \texttt{display} that gets assigned a
newly allocated \texttt{FrameDisplay} if being \texttt{null} initially.
However, in the configuration of Sunflow used by the benchmark, the value of
\texttt{display} is never \texttt{null}. Since this is the only place where the
class \texttt{FrameDisplay} is instantiated, we can conclude it is not necessary
for the execution at all.  Eventually, the \texttt{imageBegin} method is called
on \texttt{display} in the \texttt{render} method of the \texttt{BucketRenderer}
class, which for a~\texttt{FrameDisplay} transitively calls into the
\texttt{AWT} and \texttt{Swing} libraries, none of which are needed.  Yet,
a~flow-insensitive analysis is not precise enough to prove that. As it considers
neither the order of statements nor the branching structure, its pointer
assignment graph contains the spurious path \fboxsep1.5pt
\colorbox{gray!15}{$\boldsymbol{new}\; \mathtt{FrameDisplay}()$}
$\rightsquigarrow$ \colorbox{gray!15}{$\mathtt{display_{Scene}}$}
$\rightsquigarrow$ \colorbox{gray!15}{$\mathtt{display_{BucketRenderer}}$}.  On
the other hand, a~flow-sensitive analysis can cover this case (assuming it
evaluates branching conditions and tracks the \textit{nullness} of values), but
at the cost of significantly higher analysis overhead.  In \analysisName{}, the
PVPG contains a~\textit{predicate} edge
\colorbox{gray!15}{$\mathtt{display_{Scene}} == \mathtt{null}$}
$\rightsquigarrow_{pred}$ \colorbox{gray!15}{$\boldsymbol{new}\;
\mathtt{FrameDisplay}()$}, which never triggers, thus preventing
\texttt{FrameDisplay} from being considered as instantiated. This ensures that
\texttt{AWT} and \texttt{Swing} are proven unreachable.

\begin{figure}
\begin{lstlisting}[language=Java]
class SharedThreadContainer {
    Set<Thread> virtualThreads;
    public void onExit(Thread thread) {
        if (thread.isVirtual()) {
            virtualThreads.remove(thread);
        }
    }
}

class Thread {
    public boolean isVirtual() {
        return this instanceof BaseVirtualThread;
    }
}
\end{lstlisting}
\caption{A JDK motivating example.}
\label{fig:motivating-example-jdk}
\end{figure}

As a~second example, consider the program in Figure~\ref{fig:motivating-example-jdk}. 
The top part contains the method \texttt{onExit} taken from the class \texttt{SharedThreadContainer} from the \texttt{jdk.internal.vm} package. The method checks whether an exited thread is a~virtual thread and if so, the thread is removed from the set of virtual threads the class maintains. 
The actual logic of \texttt{Thread.isVirtual()} simply checks whether the thread is a~subclass of the \texttt{BaseVirtualThread} class. 

If the application does not use virtual threads, the body of the \texttt{if} statement is dead code. 
To prove that, we need an interprocedural analysis (since the condition and the type check are in different methods) that tracks both the flow of types (to prove that the type check always fails) and the flow of primitive values (to propagate the \texttt{false} value from the type check back to the caller). 
Furthermore, the analysis has to be at least partially flow-sensitive to utilize the information and consider the \texttt{remove()} call unreachable. Our analysis satisfies all these requirements and successfully removes the call. 

These examples are not artificial: the former comes from the well-established benchmark \texttt{DaCapo Sunflow} and the latter from the JDK itself. We discovered both of them when manually inspecting the results of our analysis, but they are common patterns used to dynamically configure an application. 
In our experiments, we show that handling such cases precisely has a~significant impact on the overall precision of the analysis.

\section{System Overview}
This section presents a~high-level overview of our analysis, informally describes the intuition behind it, and lists its key features. In the examples, we utilize statements \texttt{use(x)} in code snippets. Their purpose is to represent an arbitrary instruction resulting in a~\texttt{use} dependency of a~variable \texttt{x} at a~given program point.

\subsubsection*{Analysis Overview}
\label{sec:analysisOverview}

Our analysis is field-sensitive and partially flow-sensitive as explained later
in this section. The analysis starts with a~set of root methods, e.g., $\{\mathtt{main}\}$, and then processes all transitively reachable methods until
a~fixed point is reached.
Each reachable method is transformed into a \textit{predicated value propagation
graph} (PVPG) that represents the flow of both primitive values and objects.
Nodes in the PVPG are called \textit{flows}. 
Roughly, flows correspond to memory locations and instructions in the analyzed language.
We defer a more precise definition of flows to Section \ref{sec:flowsAndEdges} after
we describe the intuition behind the key features of a~PVPG.
Flows can be connected by three types
of edges: \textit{use} edges for classical value propagation; \textit{predicate}
edges used for signalling when a~given flow becomes \textit{executable}
(essentially unblocking value propagation through this flow); and
\textit{observe} edges to reflect that some flow-specific actions need to be
performed when the value of some other flow changes (e.g., when a new type appears
in a flow modeling the receiver of an invoke operation, 
the resolution and linking of a new target method might be needed).
The graphs of individual methods are connected into an interprocedural graph by
linking the actual arguments with the formal parameters and the return from the
callee back to the invocation in the caller.

\subsubsection*{Filtering Flows}
\label{sec:filteringFlows}
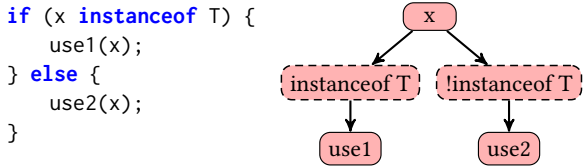
\begin{figure}
\begin{minipage}[c]{0.20\textwidth}
\begin{lstlisting}[language=Java]
if (x instanceof T) {
    use1(x);
} else {
    use2(x);
}
\end{lstlisting}
\end{minipage}
\begin{minipage}[c]{0.25\textwidth}
\begin{tikzpicture}[edge from parent/.style={draw,-latex}, level distance=3em, level 1/.style={sibling distance=6em}]
    \tikzstyle{node} = [minimum width=2em, line width=0.15mm, rectangle, rounded corners, text centered, draw=black, fill=red!30]
    \tikzstyle{filter} = [rectangle, line width=0.20mm, rounded corners, densely dashed, text centered, draw=black, fill=red!30]
    \tikzstyle{use} = [thick,->,>=stealth']

    \node [node] {\small x}
        child {node [filter, yshift=0.5em] {\small instanceof T} edge from parent [use]
            child {node [node, yshift=0.5em] {\small use1} edge from parent [use] }
        }
        child { node [filter, yshift=0.5em] {\small !instanceof T} edge from parent [use]
            child {node [node, yshift=0.5em] {\small use2} edge from parent [use] }
        };
\end{tikzpicture}
\end{minipage}
\caption{A type check example for filtering flows.}
\label{SimpleNullCheck}
\end{figure}

While \analysisName{} is mostly flow-insensitive, we maintain a certain degree
of flow-sensitivity using multiple approaches. 
First, the analysis is executed on a~base language in static single
assignment (SSA) form~\cite{SSA}, which maintains flow-sensitivity for local
variables. Second, we model conditional branches that involve type checks, null
checks, and primitive comparisons to increase the precision.
Consider the example shown in Figure~\ref{SimpleNullCheck}. 
It is obvious from the structure of the program that the value of \texttt{x} in
\texttt{use1} can only be of the type \texttt{T} (or some of its subtypes), and
similarly the value of \texttt{x} in \texttt{use2} can never be of the type
\texttt{T} (nor of any of its subtypes).
Correspondingly, the filtering flow \texttt{instanceOf T} only passes
further the type \texttt{T} (and its subtypes) for \texttt{x}, and vice versa
for the other branch.
When the \textit{value state} of a variable ends up empty in a branch, the code of that branch is proven unreachable by the analysis.

\subsubsection*{Control Flow Predicates}
\begin{figure}
\begin{minipage}[c]{0.21\textwidth}
\begin{lstlisting}[language=Java]
if (x > 10) {
    m();
} else {
    f();
}
\end{lstlisting}
\end{minipage}
\begin{minipage}[c]{0.25\textwidth}
\begin{tikzpicture}[edge from parent/.style={draw,-latex}, level distance=3em, level 1/.style={sibling distance=6em}]
    \tikzstyle{node} = [minimum width=2em, line width=0.15mm, rectangle, rounded corners, text centered, draw=black, fill=red!30]
    \tikzstyle{filter} = [rectangle, line width=0.20mm, rounded corners, densely dashed, text centered, draw=black, fill=red!30]
    \tikzstyle{disabled} = [fill=gray!30]
    \tikzstyle{use} = [thick,->,>=stealth']
    \tikzstyle{dotline} = [draw, -latex', thick, densely dashed, ->,>=stealth', arrows = {-Stealth[fill=none,length=8pt]}]

    \node (x) [node] {\small x}
        child {node [filter, yshift=0.5em] (leftCond) {\small x > 10} edge from parent [use]
            child {node [node] (leftCall) {\small invoke m()} edge from parent[draw=none]}
        }
        child { node [filter, yshift=0.5em] (rightCond) {\small x <= 10} edge from parent [use]
            child {node [node,disabled] (rightCall) {\small invoke f()} edge from parent[draw=none]}
        };
     \path (leftCond) [dotline] -- node {} (leftCall);
     \path (rightCond) [dotline] -- node {} (rightCall);

    \node (top) [left of=x, xshift=-1.0em] {42};
    \path (top) [draw,use,decorate,decoration={snake,amplitude=4pt,pre length=1pt,post length=2.5pt}] to node {} (x);
\end{tikzpicture}
\end{minipage}
\caption{A predicate example. Enabled flows are coloured in red, and disabled flows in grey.}
\label{SimplePredicate}
\end{figure}
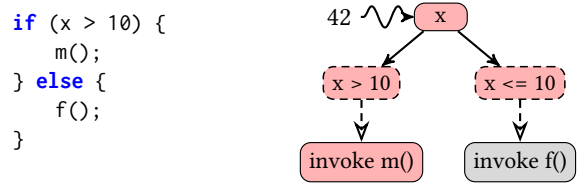
\noindent
Control flow predicates model the relationship between a~condition in a~branching instruction and the nodes within its branches. An instruction within a~\texttt{then} branch is executable iff the flow representing the condition of the branch has a~non-empty \textit{value state}, and similarly for the \texttt{else} branch. 
Contrary to the typical Andersen's points-to analysis~\cite{andersen94}, flows in our analysis only propagate values if the \textit{value state} of their predicate becomes non-empty. 
Consider the example in Figure~\ref{SimplePredicate}. 
Predicate edges are visualized using dashed lines with empty arrow heads.
If the value of \texttt{x} is known to be 42, it passes only the condition \mbox{\texttt{x > 10}} and thus only the flow representing the invoke \texttt{m()} is marked as executable. 
The condition \mbox{\texttt{x <= 10}} of the \texttt{else} branch filters \texttt{x} to \texttt{empty}, not enabling the flow representing the invoke \texttt{f()}.
In this example, a~primitive comparison acts as the~predicate. 
Similarly, any type check or null check expressed as a filtering flow is a~predicate for the beginning of its block.
When using predicates, the method \texttt{f()} is not marked as reachable and therefore not analyzed.

One might think that the above approach is covered by \emph{constant folding} commonly implemented by compilers.
However, constant folding is typically implemented within the scope of a~single method, therefore it covers only the case
when \texttt{x} is defined as a~constant locally.
In the case where \texttt{x} is not a~constant locally, e.g. it is a~method parameter, achieving the same effect using only intraprocedural optimizations is no longer possible. 

\subsubsection*{Method Invocations as Predicates}
If we can prove that a~method never returns, we can conclude that all the statements following the method invocation are unreachable. 
This happens, e.g., when a~method contains an infinite loop, or when an exception is always thrown, e.g., \texttt{Assert.fail()} and similar methods. 
In our analysis, every method invocation is a~predicate for the following statements in the block.
A method with a \texttt{void} return type still returns the predicate of the $\boldsymbol{\mathit{return}}$ instruction as an artificial value signaling whether the $\boldsymbol{\mathit{return}}$ is reachable.


\subsubsection*{Joining Values using $\phi$ Flows}
\label{sec:mergingValues}
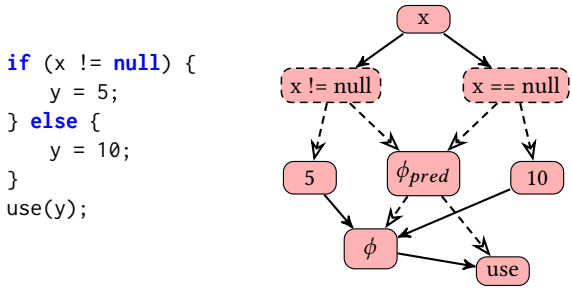
\begin{figure}
\begin{minipage}[c]{0.20\textwidth}
\begin{lstlisting}[language=Java]
if (x != null) {
    y = 5;
} else {
    y = 10;
}
use(y);
\end{lstlisting}
\end{minipage}
\begin{minipage}[c]{0.25\textwidth}
\begin{tikzpicture}[edge from parent/.style={draw,-latex}, level distance=3em, level 1/.style={sibling distance=7em}]
    \tikzstyle{node} = [minimum width=2em, line width=0.15mm, rectangle, rounded corners, text centered, draw=black, fill=red!30]
    \tikzstyle{filter} = [rectangle, line width=0.20mm, rounded corners, densely dashed, text centered, draw=black, fill=red!30]
    \tikzstyle{disabled} = [fill=gray!30]
    \tikzstyle{use} = [thick,->,>=stealth']
    \tikzstyle{dotline} = [draw, -latex', thick, densely dashed, ->,>=stealth', arrows = {-Stealth[fill=none,length=8pt]}]

    \node (x) [node] {\small x}
        child {node (leftCond) [filter, yshift=0.5em] {\small x != null} edge from parent [use]
            child {node (5) [node, xshift=-0.8em, yshift=-0.5em] {\small 5} edge from parent[draw=none]}
        }
        child { node (rightCond) [filter, yshift=0.5em] {\small x == null} edge from parent [use]
            child {node (10) [node, xshift=0.8em, yshift=-0.5em] {\small 10} edge from parent[draw=none]}
        };
    \node (PredMerge) [node, below of = x, xshift=0em, yshift=-3em] {\small $\phi_{pred}$};
    \node (Merge) [node, below of = PredMerge, yshift=0em,xshift=-2em] {\small $\phi$};
    
    \node (Use) [node, below of = Merge, xshift=5em, yshift=2em] {\small use};
    
    \draw [use] (5) -- (Merge);
    \draw [use] (10) -- (Merge);
    \draw [use] (Merge) -- (Use);

    \path (leftCond) [dotline] -- node {} (PredMerge);
    \path (rightCond) [dotline] -- node {} (PredMerge);
    
    \path (PredMerge) [dotline] -- node {} (Merge);
    \path (PredMerge) [dotline] -- node {} (Use);
    
    \path (leftCond) [dotline] -- node {} (5);
    \path (rightCond) [dotline] -- node {} (10);
\end{tikzpicture}
\end{minipage}
\caption{An example PVPG with $\phi$ flows.}
\label{mergeNode}
\end{figure}
\noindent
When multiple branches join in the control flow graph, it is necessary to join the incoming values in the PVPG. 
Consider Figure~\ref{mergeNode}. 
The value of \texttt{y} depends on which branch was taken. 
In SSA form, a~$y_3 \gets \phi(y_1,y_2)$ instruction expresses that explicitly.
In a~PVPG, a~$\phi$ flow (depicted as $\phi$) is introduced to join the two definitions of \texttt{y}. 

Each of the definitions of $y$ is guarded by their corresponding predicate, the $x \neq \texttt{null}$ flow in case of $y=5$ and the $x == \texttt{null}$ flow in case of $y=10$.
These predicates need to be joined too, producing a~predicate flow, denoted as $\phi_{pred}$, for the joined value.
Intuitively, the code after a~control flow join is executable iff the end of any of its predecessors is. 
For each control flow join, a~$\phi_{pred}$ flow is introduced, and connected via predicate edges with the latest predicates from the predecessor branches. 
This $\phi_{pred}$ becomes the new predicate in the following block, and is also the predicate of all the $\phi$ flows generated for joining values (see the \textit{pred} edge going from the $\phi_{pred}$ to $\phi$).

\subsubsection*{Abstractions for Primitive Values}
\label{sec:primAbstractions}
\begin{figure}
\begin{tikzpicture}
\node (Any) {\texttt{Any}};
\node (0) [below of=Any, yshift=0.5em] {\small \texttt{0}};
\node (1) [right of=0] {\small \texttt{1}};
\node (2) [right of=1] {\small \texttt{2 ...}};
\node (m1) [left of=0] {\small \texttt{-1}};
\node (m2) [left of=m1] {\small \texttt{... -2}};
\node (Empty) [below of=0, yshift=0.5em] {\small \texttt{Empty}};

\draw (Any) -- (0);
\draw (Any) -- (1);
\draw (Any) -- (2);
\draw (Any) -- (m1);
\draw (Any) -- (m2);
\draw (0) -- (Empty);
\draw (1) -- (Empty);
\draw (2) -- (Empty);
\draw (m1) -- (Empty);
\draw (m2) -- (Empty);
\end{tikzpicture}
\caption{The lattice $\mathbb{P}$ of primitive values.}
\label{fig:primitiveLattice}
\end{figure}
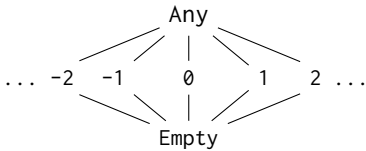
As shown in the motivating example, the analysis needs to track some primitive values such as boolean constants. 
But to scale the analysis to hundreds of thousands of methods analyzed within a few minutes, it is necessary to provide sufficient abstractions. 
Therefore, we use the simple lattice $\mathbb{P}$ depicted in Figure~\ref{fig:primitiveLattice} to model primitive values.
We do not attempt to model intervals or sets, only concrete values, \texttt{Empty}, and \texttt{Any}. 
The join of two different constant values results immediately in \texttt{Any}. 
The boolean constants \texttt{true} and \texttt{false} are modelled as constants 1 and 0, respectively.
Similarly, we do not explicitly model arithmetic operations in the PVPG. 
Instead, a~flow is inserted into the graph that always produces the value \texttt{Any}.

\dk{
\subsubsection*{Abstractions for Objects}
\label{sec:abstractionsForObjects}
\analysisName{} is designed to work with any lattice-based representation for
objects as long as type checks, null checks, and virtual method resolution can
be applied on the elements of the lattice.  Null references are handled as a
special type that can be part of any value state.  In the implementation of
\analysisName{} evaluated in this paper, we have chosen to represent objects by their types only.  This
\textit{type abstraction} was shown useful by prior work when embedding
the analysis in a~compiler~\cite{Wimmer24}.  On top of that, the type
abstraction (or a similar mechanism) is in some cases even necessary to preserve
soundness for Java: For example, allocation-site information is not available when interfacing with C code via JNI where new Java objects can be created in C or passed back and forth between Java and C, or object fields can be
queried in C and the results then passed back into Java.

However, we argue that our technique is applicable to both type-based and object-based points-to analysis.
The decision to use the type abstraction is an implementation detail as the same analysis can be executed using a subset lattice based on allocation sites, possibly even extended with allocation context for context-sensitive analysis. 
}

\section{Predicated Value Propagation Graphs}
\label{sec:PVPG}
In this section, we describe the structure of PVPGs and the rules for value propagation over them. 
For space reasons, we focus on a~high-level description.
The full formalism can be found in Appendices~\ref{sec:app:pvpg} and \ref{sec:valuepropagationrules}.
The input to the analysis is a~Java-like managed base language in static single assignment form.

\subsubsection*{Structure of a PVPG}
\label{sec:flowsAndEdges}

A~PVPG models the flow of both primitive values and objects interprocedurally. Vertices in a~PVPG are called \textbf{flows} to clearly distinguish them from the \textbf{instructions} in the base language.
The flows represent values of method parameters, variables, and fields read or written by a particular instruction (each instruction reading or writing to a variable produces a fresh flow); method calls, which also represent the returned value in the caller; and values to be returned from a method call back to the caller.
Further, flows represent values of various conditions present in the code (including their negated and inverted versions as explained later); values resulting from joining the values of other flows (results of $\phi$  instructions); $\phi_{pred}$ flows for joining predicates; and, finally, the always enabled predicate \predEnabled{}.
Each flow keeps a reference to the underlying base language element it represents.
The analysis computes the \textit{value state} of each flow representing a~conservative overapproximation of the values that can be assigned to the given flow during the runtime of the program. 

Flows can be connected via three types of \textbf{edges} described below.

A \textit{use} edge represents a~\textit{def-use} dependency between flows. 
If there is a~\textit{use} edge between flows \texttt{s} and \texttt{t}, denoted by \mbox{$\mathtt{s}\rightsquigarrow_{use}\mathtt{t}$}, the \textit{value state} of \texttt{t} has to be at least as big as the \textit{value state} of \texttt{s} in terms of the underlying lattice provided that $s$ is executable and no filtering is applied (the semantics of filtering is explained later).

A \textit{predicate} edge represents a~control-flow dependency. A~\textit{predicate} edge between flows \texttt{s} and \texttt{t}, denoted by \mbox{$\mathtt{s}\rightsquigarrow_{pred}\mathtt{t}$}, defines that if \texttt{s} is executable and has a~non-empty \textit{value state}, \texttt{t} is also executable. 
Every flow is the target of an incoming \textit{predicate} edge, apart from \predEnabled{}, which is always enabled and therefore does not need an incoming edge.
A~$\phi_{pred}$ flow has multiple incoming \textit{predicate} edges. 
In that case, it suffices that any of its predicates has a~non-empty state to enable the execution of the target flow.

An \textit{observe} edge represents an additional flow-specific dependency. An \textit{observe} edge between flows \texttt{s} and \texttt{t}, denoted as \mbox{$\mathtt{s}\rightsquigarrow_{obs}\mathtt{t}$}, defines that \texttt{t} has access to the \textit{value state} of \texttt{s}, and \texttt{t} is notified when the state changes to perform some flow-specific task such as a~field update or method resolution and linking (discussed more below).

\subsubsection*{Creating a PVPG}
This paragraph describes how the PVPG can be created for a~given method.
A~formalization of the algorithm is presented in Appendix~\ref{sec:app:pvpg}.
Basic blocks of the method are processed in reverse postorder,
and the instructions within each basic block are processed sequentially top to bottom.
During the traversal, flows are created for the encountered instructions and memory locations.

The traversal maintains a state for each basic block consisting of:
(1) A~mapping from variables to previously created flows, which is used
to connect flows with their dependencies, e.g. to establish a \textit{use} edge between the flow representing \texttt{x} and the subsequently created filtering flow \texttt{x != 0}, and
(2) a~reference denoted as \textit{pred} that is continuously updated to refer to the last encountered predicate.
The predicate referenced by \textit{pred} is used to establish predicate edges.
At the beginning of a~method, when no suitable flow is available yet, the special \predEnabled{} flow is used, which is always enabled, i.e. the initial flows inside methods are always enabled.

When processing a~branching instruction, the branching condition is handled separately for the \texttt{then} and \texttt{else} branch. 
For the \texttt{then} branch, the condition is used as is; for the \texttt{else} branch, it has to be negated, e.g. \texttt{x < 10} becomes \texttt{x >= 10}. 
These conditions are used to filter the values of their tested variables within the scope of the successor branches, which is done using the filtering flows described in Section \ref{sec:filteringFlows}.
Inside each branch, the following happens.
If the condition tests a~single variable only, for example a~\textit{nullcheck} or a~\textit{typecheck}, a~single filtering flow is created. 
If the condition is a~comparison of two variables, e.g. \texttt{x < y}, two flows are created, each representing one variable\footnote{Note that, here, we assume a~base language with conditions involving at most two variables as compound expressions can be broken into chains of simpler ones without loss of generality.}.
The filtering flows serve as new definitions for the variables they test.
In our example, we have a~new flow for \texttt{x} whose value is less than \texttt{y}, and a~new flow for \texttt{y}, whose value is greater than \texttt{x}.


\subsubsection*{Running Example}
\label{sec:runningExamplePVPG}
\begin{figure}
\centering
\scalebox{0.85}{
\begin{tikzpicture}[edge from parent/.style={draw,-latex}, level distance=1em, level 1/.style={sibling distance=8em}]
    \tikzstyle{node} = [minimum width=1.5em, line width=0.15mm, rectangle, rounded corners, text centered, draw=black, fill=red!30]
    \tikzstyle{filter} = [rectangle, line width=0.20mm, rounded corners, densely dashed, text centered, draw=black, fill=red!30]
    \tikzstyle{disabled} = [fill=gray!30]
    \tikzstyle{use} = [thick,->,>=stealth']
    \tikzstyle{dotline} = [draw, -latex', thick, densely dashed, ->,>=stealth', arrows = {-Stealth[fill=none,length=8pt]}]
    \tikzstyle{observed} = [draw, anchor=west, thick, -latex',densely dotted, ->]

    \node (eq) [filter, yshift=2em, minimum width=1.5em,minimum height=1.35em, disabled] {\small $\neq$} [grow'=up] 
        child { 
            node (isVirtual) [node,disabled, xshift=-1.5em, yshift=1.3em] {\small Invoke isVirtual()} edge from parent[draw=none] 
        }
        child { 
            node (zeroL) [node,disabled, xshift=-1.5em, yshift=1.3em] {\small 0} edge from parent[draw=none] 
        }
    ;

    \node (eq2) [xshift=0em, yshift=0.0em] {};
    
    \node (paramThread) [node,disabled, left of = isVirtual, xshift=-4em, yshift=1.4em] {\small p\textsubscript{thread}};
    \node (thisL) [node, disabled, left of = eq2, xshift=-6em, yshift=1.9em] {\small p\textsubscript{this}};
    \node (loadField) [node, disabled, below of = thisL, xshift=0em, yshift=-0.5em] {\small LoadField virtualThreads};
    \node (remove) [node, disabled, below of = eq, xshift=-4.3em, yshift=-4em] {\small Invoke remove()};

    \path ($(isVirtual.south) + (1,0)$) [draw,use] to[out=-90,in=135] node [xshift=0.2,yshift=0.35em] {} (eq);
    \draw [use,dotline] (isVirtual) to[out=0,in=90] (eq);
    
    \path ($(zeroL.south) + (-0.25,0)$) [observed, anchor=east] to[out=-90,in=45] node [xshift=0.15em, yshift=0.20em] {} (eq);
    \path ($(paramThread.south) + (0.0,0)$) [observed, anchor = west] to[out=-90,in=180] node [xshift=-0.0] {} ($(isVirtual.west) + (-0.0,0)$);
    \path (thisL) [observed, anchor=east] -- node [xshift=0.2em] {} (loadField);

    \path (loadField.south) [observed, anchor=east] to[out=-90,in=180] node [xshift=0.5em,yshift=0.6em] {} (remove);

    \path (eq.south) [dotline, anchor = north] to [out=-135,in=45] node {} ($(loadField.north east) + (0,0)$);
    \path (eq.south) [dotline] to[out=-90,in=45] node {} (remove.north east);

    \node (thisR) [node, disabled, right of = zeroL, xshift=2.0em, yshift=-2em] {\small p\textsubscript{this}}
        child {
            node (is) [filter, disabled, yshift=-1em] {\small instanceof T} edge from parent[draw=none] 
        }
        child{
            node (notIs) [filter, disabled, yshift=-1em] {\small !instanceof T} edge from parent[draw=none] 
        }
    ;

    \draw [use] (thisR.west) to[out=180,in=90] (is);
    \draw [use] (thisR.east) to[out=0,in=90] (notIs);
    
    \node (1R) [node, disabled, below of = is, yshift=-0.6em] {\small 1};
    \node (0R) [node, disabled, below of = notIs, yshift=-0.6em] {\small 0};

    \path (is) [dotline, anchor = east] -- node {} (1R);
    \path (notIs) [dotline] -- node {} (0R);

    \node (mergePred) [node,disabled, below of = thisR, yshift=-1.8em] {\small $\phi_{pred}$};
    \path (is) [dotline] -- node {} (mergePred);
    \path (notIs) [dotline, anchor = east] -- node {} (mergePred);

    \node (merge) [node, yshift=0.1em, disabled, below of = mergePred] {\small $\phi$};
    \path (mergePred) [dotline] -- node {} (merge);
    \draw [use] (1R.south) to[out=-90,in=180] (merge.west);
    \draw [use] (0R.south) to[out=-90,in=0] (merge.east);
    \node (ret) [node, disabled, below of = merge, yshift=0.3em] {\small Return};
    \draw [use] (merge) -- (ret);
    \path (mergePred) [dotline] to[out=-25,in=45] node {} (ret.east);
\end{tikzpicture}
}
\caption{The PVPG for the methods \texttt{onExit} (on the left) and \texttt{isVirtual} (on the right) from the JDK motivating example in Figure~\ref{fig:motivating-example-jdk}. Full lines represent \textit{use} edges, dashed lines represent \textit{predicate} edges, and dotted lines represent \textit{observe} edges. In the instanceof check, T stands for \texttt{BaseVirtualThread}.
Note that the two p\textsubscript{this} flows each represent the implicit \texttt{this} parameter in their corresponding method.
}
\label{fig:runningExampleGraph}
\end{figure}
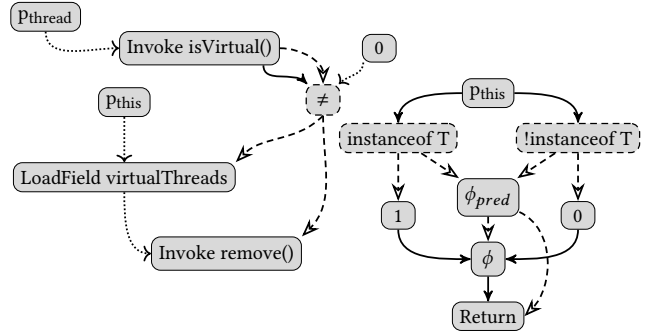
Figure~\ref{fig:runningExampleGraph} shows the PVPG for the methods \texttt{onExit} (on the left) and \texttt{isVirtual} (on the right) from the JDK motivating example in Figure~\ref{fig:motivating-example-jdk}. 
To make the graph more compact, we omit the always-enabled predicate \predEnabled{}.
A~flow that does not have any incoming \textit{predicate} edge in the graph is predicated by \predEnabled{}.
The condition that invokes \texttt{isVirtual()} and that guards the \texttt{then} branch containing the Invoke \texttt{remove()} leads to the filtering flow denoted as '$\neq$'.
Recall that boolean values are represented as integers, thus the condition becomes \mbox{\texttt{if (thread.isVirtual() != 0)}}, which is encoded explicitly in the graph.
The flow representing the Invoke \texttt{isVirtual()} is the predicate of '$\neq$' (Indeed, the method must be first invoked and only then the test may be executed),  which in turn is the predicate of the flow representing the load field \texttt{virtualThreads} and the flow representing the  \texttt{remove()} invocation.

Furthermore, note that p\textsubscript{thread} is connected via \textit{observe} edge with the Invoke \texttt{isVirtual()}, because the invocation may need to link new call targets every time a new type appears in the \textit{value state} of p\textsubscript{thread}.
The same applies to the chain of \textit{observe} edges from p\textsubscript{this} through the load field of \texttt{virtualThreads} to the Invoke \texttt{remove()}. 
Every time a~new type appears in the \textit{value state} of p\textsubscript{this}, a~new field may be connected with the load field instruction, which in turn may add more types to the \textit{value state} of the load field, possibly triggering a new method linking for the Invoke \texttt{remove()}. 

The filtering flow '$\neq$' is connected via a~\textit{use} edge with the Invoke \texttt{isVirtual()}, whose values it filters and propagates. 
Further, it also has an \textit{observe} edge from the constant zero, because it needs to access its \textit{value state} to perform the filtering (in this case, the value is constant, but it does not have to be in general). 

In the \texttt{isVirtual()} method (on the right), the type check leads to two filtering flows, each of which is the predicate for their corresponding returned values.

\subsubsection*{Value propagation through PVPGs}
This section describes the rules for value propagation in a~PVPG.
A~formalization is given in Appendix~\ref{sec:valuepropagationrules}.
Each flow maintains a~\textit{value state} representing the~set of values that can be assigned to the corresponding base language element at runtime.
Primitive values are modelled using the lattice $\mathbb{P}$ discussed in Section~\ref{sec:primAbstractions}. 
Objects are modelled using a~subset lattice over types.
Following the work of Wimmer et al.~\cite{Wimmer24}, we do not distinguish individual allocation sites to improve scalability.  

Each flow, except from \predEnabled{}, is initially disabled, not propagating any values.
At the beginning of the analysis, \predEnabled{} enables all the flows to which it is directly connected via a \textit{predicate} edge.
Once a flow becomes enabled, it propagates its own \textit{value state} along the \textit{use} edges.
If a flow has multiple incoming \textit{use} edges, its \textit{value state} is the join of all the incoming values.
Additionally, filtering flows filter their incoming values based on their conditions, e.g. a~type check allows only subtypes to flow through it, x < 10 allows only values smaller than 10 to flow on, etc.
Once a \textit{value state} of any enabled flow becomes non-empty, it enables all the flows to which it is connected via \textit{predicate} edges,
possibly triggering more value propagation.

Method invocations are handled by \textit{linking}, i.e. creating \textit{use} edges from the arguments in the caller to the formal parameters of the callee, and from the return in the callee back to the invoke flow in the caller (the invoke flow also represents the returned value).
For virtual invokes, the linking is done for every possible target method obtained by inspecting the \textit{value state r} of the receiver (with which the invoke is connected via an~\textit{observe} edge) and performing \textit{method resolution}\footnote{The virtual method resolution is done as defined by the JVM specification~\cite{JVMSpec21}.} for each type $t \in r$.

Since the elements used in \textit{value states} form a~lattice with a~finite height and all our filtering and join operators are monotone, the analysis is guaranteed to eventually reach a~fixed point and terminate.

\subsubsection*{Running Example}
\label{sec:runnigExampleAfterAnalysis}
\begin{figure}
\centering
\scalebox{0.82}{
\begin{tikzpicture}[edge from parent/.style={draw,-latex}, level distance=1em, level 1/.style={sibling distance=8em}]
    \tikzstyle{node} = [minimum width=1.5em, line width=0.15mm, rectangle, rounded corners, text centered, draw=black, fill=red!30]
    \tikzstyle{filter} = [rectangle, line width=0.20mm, rounded corners, densely dashed, text centered, draw=black, fill=red!30]
    \tikzstyle{disabled} = [fill=gray!30]
    \tikzstyle{use} = [thick,->,>=stealth']
    \tikzstyle{dotline} = [draw, -latex', thick, densely dashed, ->,>=stealth', arrows = {-Stealth[fill=none,length=8pt]}]
    \tikzstyle{observed} = [draw, anchor=west, thick, -latex',densely dotted, ->]

    \node (eq) [filter, yshift=2em,minimum width=1.5em,,minimum height=1.35em] {\small $\neq$} [grow'=up] 
        child { 
            node (isVirtual) [node, xshift=-1.5em, yshift=1.3em] {\small Invoke isVirtual()} edge from parent[draw=none] 
        }
        child { 
            node (zeroL) [node, xshift=-1.5em, yshift=1.3em] {\small 0} edge from parent[draw=none] 
        }
    ;

    \node (eq2) [xshift=0em, yshift=0.0em] {};
    
    \node (paramThread) [node, left of = isVirtual, xshift=-4em, yshift=1.4em] {\small p\textsubscript{thread}};
    \node (thisL) [node, left of = eq2, xshift=-6em, yshift=1.9em] {\small p\textsubscript{this}};
    \node (loadField) [node, disabled, below of = thisL, xshift=0em, yshift=-0.5em] {\small LoadField virtualThreads};
    \node (remove) [node, disabled, below of = eq, xshift=-4.3em, yshift=-4em] {\small Invoke remove()};

    \path ($(isVirtual.south) + (1,0)$) [draw,use] to[out=-90,in=135] node [xshift=0.2,yshift=0.35em] {} (eq);
    \draw [use,dotline] (isVirtual) to[out=0,in=90] (eq);
    
    \path ($(zeroL.south) + (-0.25,0)$) [observed, anchor=east] to[out=-90,in=45] node [xshift=0.15em, yshift=0.20em] {} (eq);
    \path ($(paramThread.south) + (0.0,0)$) [observed, anchor = west] to[out=-90,in=180] node [xshift=-0.0] {} ($(isVirtual.west) + (-0.0,0)$);
    \path (thisL) [observed, anchor=east] -- node [xshift=0.2em] {} (loadField);

    \path (loadField.south) [observed, anchor=east] to[out=-90,in=180] node [xshift=0.5em,yshift=0.6em] {} (remove);

    \path (eq.south) [dotline, anchor = north] to [out=-135,in=45] node {} ($(loadField.north east) + (0,0)$);
    \path (eq.south) [dotline] to[out=-90,in=45] node {} (remove.north east);

    \node (thisR) [node, right of = zeroL, xshift=1.8em, yshift=-2em] {\small p\textsubscript{this}}
        child {
            node (is) [filter, yshift=-1em] {\small instanceof T} edge from parent[draw=none] 
        }
        child{
            node (notIs) [filter, yshift=-1em] {\small !instanceof T} edge from parent[draw=none] 
        }
    ;

    \draw [use] (thisR.west) to[out=180,in=90] (is);
    \draw [use] (thisR.east) to[out=0,in=90] (notIs);
    
    \node (1R) [node, disabled, below of = is, yshift=-0.6em] {\small 1};
    \node (0R) [node, below of = notIs, yshift=-0.6em] {\small 0};

    \path (is) [dotline, anchor = east] -- node {} (1R);
    \path (notIs) [dotline] -- node {} (0R);

    \node (mergePred) [node, below of = thisR, yshift=-1.8em] {\small $\phi_{pred}$};
    \path (is) [dotline] -- node {} (mergePred);
    \path (notIs) [dotline, anchor = east] -- node {} (mergePred);

    \node (merge) [node, yshift=0.1em, below of = mergePred] {\small $\phi$};
    \path (mergePred) [dotline] -- node {} (merge);
    \draw [use] (1R.south) to[out=-90,in=180] (merge.west);
    \draw [use] (0R.south) to[out=-90,in=0] (merge.east);
    \node (ret) [node, below of = merge, yshift=0.3em] {\small Return};
    \draw [use] (merge) -- (ret);
    \path (mergePred) [dotline] to[out=-25,in=45] node {} (ret.east);

    \path ($(paramThread.north east) + (0,0)$) [draw,use,anchor = south] to[out=15,in=145] node [xshift=0em,yshift=0.0em] {} ($(thisR.north) + (0.0,0)$);
    
    \path ($(ret.west)$) [draw,use,anchor=north]  to[out=170,in=315] node [xshift=1.5em,yshift=-4.0em] {\small $\mathit{VS(Return)} = \{0\}$} ($(isVirtual.south) + (0.4,0)$);
    
    \node (input) [above of = paramThread, xshift=3.5em, yshift=0.2em] {\small $\mathtt{VirtualThread} \notin \mathit{VS(p_{thread})}$};
    \path ($(input.south) + (-1.15,0)$) [draw, use, anchor = west]  to[out=-90,in=90] node [yshift=0.0em] {} ($(paramThread.north) + (0,0)$);
\end{tikzpicture}
}
\caption{The PVPG for the methods \texttt{onExit} (on the left) and \texttt{isVirtual} (on the right) from the JDK motivating example in Figure~\ref{fig:motivating-example-jdk} after the analysis reaches a~fixed-point. Full lines represent \textit{use} edges, dashed lines represent \textit{predicate} edges, and dotted lines represent \textit{observe} edges. Enabled flows are coloured in red, and disabled flows in grey. In the instanceof check, T stands for \texttt{BaseVirtualThread}. 
The method \texttt{onExit} is linked from some caller, but its \texttt{p\textsubscript{thread}} never receives any VirtualThread (\textit{VS} stands for \textit{value state}). Note that the two p\textsubscript{this} flows each represent the implicit \texttt{this} parameter in their corresponding method.}
\label{fig:runningExampleGraphAfterAnalysis}
\end{figure}
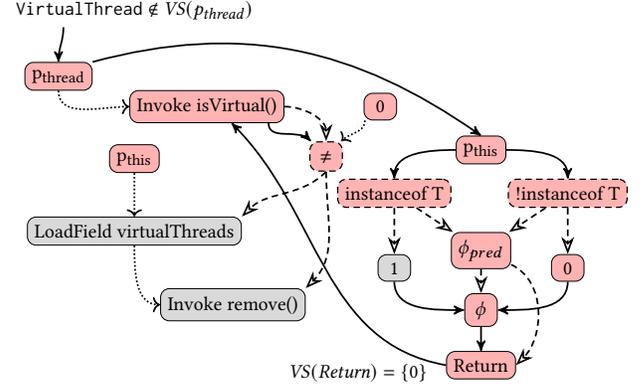

Figure~\ref{fig:runningExampleGraphAfterAnalysis} shows the state of the PVPG for the JDK motivating example after the analysis reaches the~fixed point.
The following steps produce this state:
First, the method \texttt{onExit()} is linked from some already reachable method \texttt{m}.
Then, the PVPG of \texttt{onExit()} is created and its \texttt{p\textsubscript{thread}} is linked with the flow in \texttt{m} representing the corresponding argument. 
The flows \texttt{p\textsubscript{thread}}, Invoke \texttt{isVirtual()}, the flow representing the constant 0, and \texttt{p\textsubscript{this}} from the \texttt{onExit()} method get immediately enabled because they are guarded by \predEnabled{}. 

Once the \textit{value state} of \texttt{p\textsubscript{thread}} is non-empty, the Invoke \texttt{isVirtual()} is notified via its \textit{observe} edge, which leads to a~creation of the PVPG for the \texttt{isVirtual()} method and to linking the callee and the caller.
During the linking, \texttt{p\textsubscript{thread}} in \texttt{onExit()} is linked with \texttt{p\textsubscript{this}} in \texttt{isVirtual()}, and the \texttt{Return} from \texttt{isVirtual()} is linked with the Invoke \texttt{isVirtual()} in \texttt{onExit()}, since the \textit{value state} of the flow representing the invocation represents the returned value in the caller.
So far, the \textit{value state} of the \texttt{Return} in \texttt{isVirtual()} is still empty.
In \texttt{isVirtual()}, the only enabled flows after the creation are \texttt{p\textsubscript{this}} and both filtering flows representing the type check.
Assume that no virtual thread is created in the analyzed application, so the \textit{value state} of \texttt{p\textsubscript{thread}} only contains non-virtual thread types.
Consequently, only the filtering flow for the \texttt{else} branch in \texttt{isVirtual()} receives a~non-empty \textit{value state} after filtering, which enables the flow holding the constant $0$. 
This value is propagated back into \texttt{onExit()} through the $\phi$ and the \texttt{Return} flows.

The \textit{value state} of the Invoke \texttt{isVirtual()} is now non-empty and enables the filtering flow '$\neq$'. 
However, the \textit{value state} of the Invoke \texttt{isVirtual()} contains zero only, which is filtered out by $\neq$ with another zero, so the \textit{value state} of '$\neq$' remains empty, never enabling the rest of the flows in \texttt{onExit}.
Therefore, the Invoke \texttt{remove()} is not enabled, and consequently, the \texttt{remove()} method is not processed.

\section{Implementation Details}
This section presents various details that are more tied to our specific implementation than the general algorithm presented above. 
We also discuss how we handle dynamic parts of Java including Reflection and JNI. 

\subsubsection*{Handling Exceptions}
\label{sec:exceptions}
While it is possible to track the propagation of exceptions from callees back into callers, we have not observed enough precision improvement in our experiments to justify the overhead. 
Therefore, to improve the scalability of the analysis, we assume that any instantiated exception can flow out of any exception handler with matching types. 

Exception handling provides another use case for \textit{method invokes as predicates}. 
Certain methods may never return and instead always throw exceptions. 
Such a~behaviour can either be intentional, e.g., \texttt{Assert.fail()} and similar methods, or accidental, for example when the arguments passed into a~method are invalid, e.g., \texttt{null} is passed as an argument that should be \texttt{non-null}. 
In such cases, \analysisName{} can prove that the code after the method invocation is unreachable. 


\subsubsection*{Boolean Values}

Consistent with the JVM Specification~\cite{JVMSpec21}, we model \texttt{boolean} values \texttt{true} and \texttt{false} as 1 and 0, respectively. 
Consequently, the type \texttt{boolean} does not exist from the point of view of the analysis. 
Recall that predicate edges are always triggered when the \textit{value state} becomes non-empty, which is different from the \textit{value state} becoming 0, i.e. \texttt{false}.
When a \textit{value state} becomes 0, it is non-empty, so the value propagation is triggered.
Figure~\ref{fig:runningExampleGraph} of the running example showed how conditions are converted to comparisons with the constants 1 and 0.

\subsubsection*{Reflection, JNI, Unsafe}

The Java Virtual Machine Specification~\cite{JVMSpec21} contains many dynamic features that are hard to analyse statically~\cite{Landman17}, for example, the Reflection API, JNI, and \texttt{Unsafe}. 
While soundness is in practice sometimes replaced with soundiness~\cite{Livshits15}, our analysis is meant to be embedded in a~compiler, thus we have to handle all of these dynamic features soundly. 
Otherwise, incorrect code would be generated. 
On the other hand, allowing any method to be called in a~way opaque to the analysis defeats the purpose of the analysis, as everything would be reachable.

Therefore, we require a~configuration file specifying which methods and fields
are accessed via Reflection or JNI. \dk{We inherit this design decision from
GraalVM Native Image~\cite{Wimmer19}, upon which the prototype of
\analysisName{} is implemented. One might argue that such an approach could
lead to unsound results if incomplete configuration is provided. However, we
believe that this approach is still sound with respect to the semantics of GraalVM
Native Image, which relies on these configuration files anyway. We do not claim
any contributions in the area of reflection handling, and \analysisName{} could be easily adjusted to use a different policy.}

To help creating the configuration, we provide a~tool similar to~\cite{Bodden11}: a~tracing agent that monitors the application at runtime.
For methods designated as invoked via Reflection or JNI, we mark them as \textit{root methods} and assume that the \textit{value states} of their parameters can contain any instantiated subtype of their declared type.
Similarly, the \textit{value state} of any field accessed via Reflection or JNI can contain any instantiated subtype of its declared type.

For \texttt{Unsafe}, we conservatively assume that any input of any unsafe write or normal write into an unsafe-accessed field can flow out of any unsafe load.

\section{Evaluation}
\label{section:Evaluation}
We have implemented our approach in the Native Image~\cite{Wimmer19} component of GraalVM, which produces standalone binaries for Java applications that contain the application along with all its dependencies, as well as the necessary runtime components such as the garbage collector. 
We base our branch on a~recent commit in \texttt{master}\footnote{Commit id: \href{https://github.com/oracle/graal/commit/d82a349401e5cfeb7d4523dd16810bfaa28b60b7}{$d82a349401e5cfeb7d4523dd16810bfaa28b60b7$}.}. 
In the experiments, which were done on top of Oracle GraalVM, we compare our
analysis (depicted as \analysisName{}), with the points-to analysis (depicted as
PTA) used by default in Native Image (type-based flow-insensitive
\dk{context-insensitive} analysis)~\cite{Wimmer24}. \dk{That is the
configuration shown in~\cite{Wimmer24} to be the most suitable when the analysis results are used for compilation, i.e., when a~fully sound analysis is needed.}

\dk{
A comparison between PTA and other commonly used call graph construction
algorithms, namely, Rapid Type Analysis (RTA)~\cite{fast_analysis_cpp} and
Variable Type Analysis (VTA)~\cite{practical_method_resolution_java}, was
already presented in~\cite{Wimmer24} (their \textit{context-insensitive without saturation} configuration is essentially a VTA). The remaining typical call graph construction algorithm, Class Hierarchy Analysis (CHA)~\cite{Dean95}, is expected to be even less precise than RTA. Since the precision of RTA is already too low, a CHA is not implemented in GraalVM Native Image at the moment. 
}

All the experiments were executed with Java~24\footnote{We chose to present the data using a~cutting edge release, but we have also experimented with Java 21, producing similar results.}.
We run all the benchmarks 10~times and report the average of the runs. 
The benchmarks were executed on a dual-socket Intel Xeon E5-2630 v3 running at 2.40~GHz with 8~physical / 16~logical cores per socket, 128~GByte main memory, running Oracle Linux Server release~7.3. 
The benchmark execution was pinned to one of the two CPUs, and TurboBoost was disabled to avoid instability. 
The number of threads used by the analysis was set to~16.
We use a~server configuration for benchmarking because that allows us to provide stable and reproducible numbers. However, we deliberately selected an old configuration similar to a~current developer laptop.

We use three benchmark suites: DaCapo~9.12~\cite{Blackburn06}, Renaissance~0.15.0~\cite{renaissance}, and \emph{microservices}. 
\textit{DaCapo} is a~benchmark suite that consists of client-side Java benchmarks, trying to exercise the complex interactions between the architecture, compiler, virtual machine and running application. 
\textit{Renaissance} is a~benchmark suite that consists of real-world, concurrent, and object-oriented workloads that exercise various concurrency primitives of the JVM. 
In both \textit{Renaissance} and \textit{DaCapo}, we use a~subset of the benchmark suite because some benchmarks are not compatible with the Native Image Ahead-of-Time compilation. 
However, we decided to include the benchmarks \textit{als}, \textit{chi-square}, \textit{dec-tree}, and \textit{log-regression} from \textit{Renaissance}, which are not yet fully supported in Native Image. 
In particular, their analysis already finishes successfully, but the compiled application does not run because class definitions at runtime are used, which are not yet fully supported. 

\textit{Microservices} is our own set of microservice applications. 
We use the three most frequently used modern frameworks for Java web services in our evaluation: Spring, Micronaut, and Quarkus. 
In order not to be biased towards a single framework, we selected multiple representative applications for each of them. 
All the applications perform different functions. 
Under no circumstances can our evaluation be seen as a~comparison and ranking of the frameworks against each other. 

In particular, the microservices benchmark suite consists of the following applications:
\begin{itemize}
    \item \emph{Micronaut Helloworld}: A~helloworld application written with the Micronaut framework.
    
    \item \emph{Micronaut MuShop Order, Payment, and User}~\cite{MuShop}: Three microservices of a~large demo application using the Micronaut framework.

    \item \emph{Quarkus Helloworld}: A~helloworld application written with the Quarkus framework.

    \item \emph{Quarkus Registry}~\cite{QuarkusRegistry}: A large real-world application using the Quarkus framework. It is used to host the Quarkus extension registry.

    \item \emph{Quarkus Tika}: A~demo application written with the Quarkus framework.
    
    \item \emph{Spring Helloworld}: A~helloworld application written using the Spring framework.
    
    \item \emph{Spring PetClinic}~\cite{PetClinic}: A large demo application for the Spring framework.
\end{itemize}

For each benchmark, we collect the following metrics:
\begin{itemize}
    \item \emph{Analysis Time}: The time it takes to run the analysis, measured in milliseconds.
    \item \emph{Reachable Methods}: The number of methods marked reachable by the analysis.
    \item \emph{Counter Metrics}: We count specific instructions in all reachable methods that cannot be removed or simplified using the results of the analysis. In particular, we count three types of branching instructions: \emph{Type Checks}, \emph{Null Checks}, \emph{Primitive Checks}, and the number of virtual method calls that could not be devirtualized, denoted as \emph{PolyCalls}. 
    \dk{\item \emph{Total Time}: The time it takes to run the whole Native Image compilation, measured in milliseconds.}
    \dk{\item \emph{Binary Size}: The size of the resulting binary file, measured in MB.}
\end{itemize}
\dk{In the discussions below, we focus mainly on the first four metrics, which we consider \textit{analysis-oriented}.  
The \textit{total time} and \textit{binary size} are added to demonstrate the effect on the whole Native Image compilation.
}

\begin{table*}
\centering
\caption{Results for all bench suites. For all metrics, lower is better. The best result in \textit{reachable methods} for each bench suite is in \colorbox{myyellow!100}{yellow}, the worst is in \colorbox{mygray!100}{grey}. Even for the \colorbox{mygray!100}{grey} rows, \analysisName{} still improves over the baseline in all metrics apart from \textit{analysis time}.}
\label{tab:detailedtable}
\includegraphics[width=\textwidth]{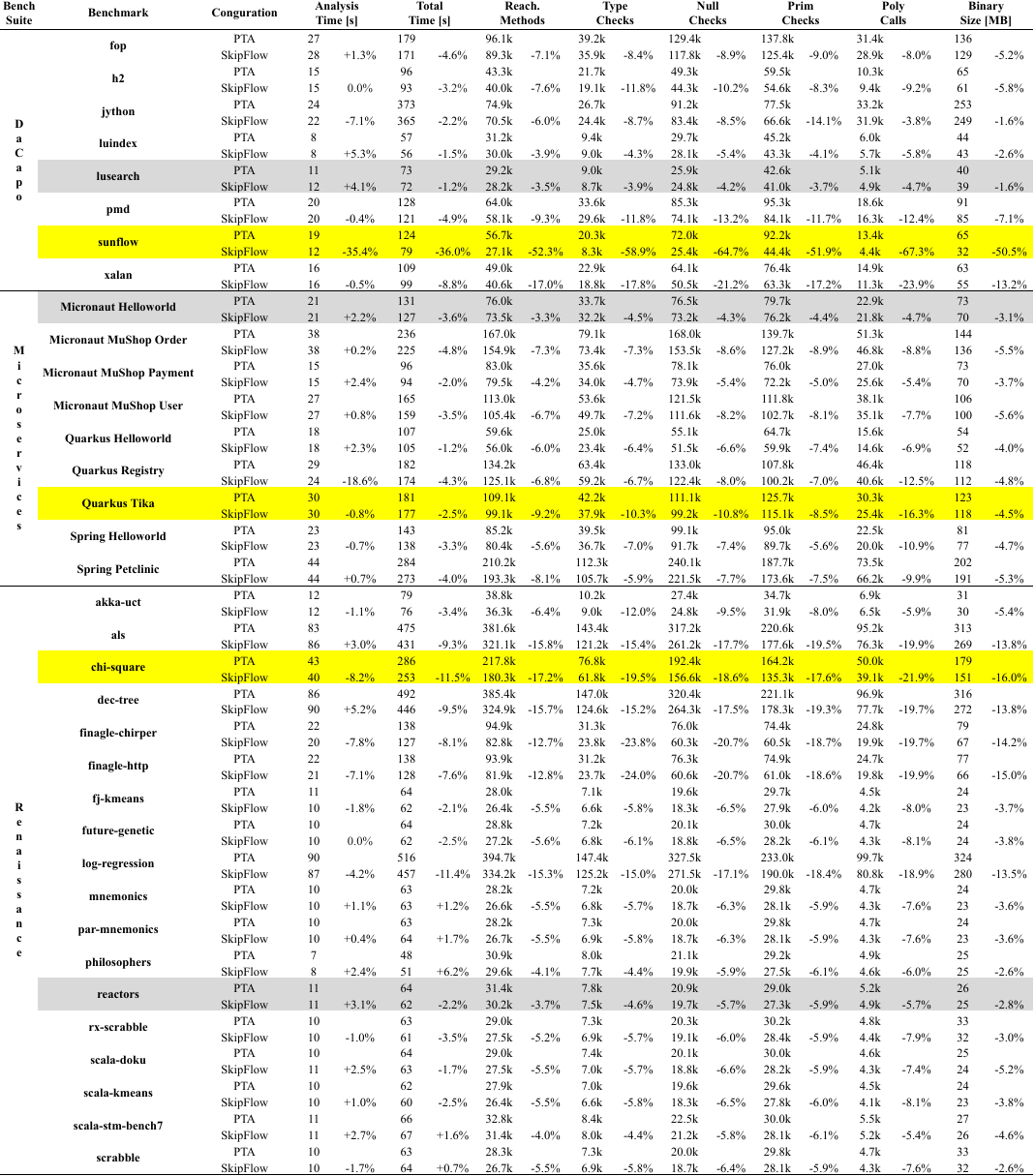}
\end{table*}

\subsubsection*{Detailed Results}
Table~\ref{tab:detailedtable} presents the results for all the benchmarks. 
Note that due to the rounding to seconds, the analysis time in the table sometimes looks identical, but the actual values in terms of milliseconds are still slightly different.
This can be observed for example for the \textit{DaCapo luindex} benchmark.

The first block in the table presents the \textit{DaCapo} benchmarks. 
The most interesting example in this category is the \textit{Sunflow} benchmark, in which the size of the application is reduced by \NUMBER{52.3\%}. 
The reason for such a~radical reduction is that fact that our analysis successfully removed the \texttt{AWT} and \texttt{Swing} GUI libraries, as explained in the motivating example in Section~\ref{sec:motivatingExample}.
The least impacted benchmark is \texttt{lusearch}, reduced by \NUMBER{3.5\%}. 
\texttt{lusearch} is also the smallest benchmark in this category, thus offering fewer program points where the precision can be improved.
On average, the number of reachable methods is reduced by \NUMBER{13.3\%}. 
The counter metrics follow a similar trend. 

The most prominent case for \textit{Microservices} is \texttt{Quarkus Tika}, whose size is reduced by \NUMBER{9.2\%}, followed by the biggest benchmark in this category, \texttt{Spring Petclinic}, which is reduced by \NUMBER{8.1\%}.
The least impacted benchmark is \texttt{Micronaut Helloworld}, reduced by \NUMBER{3.3\%}, again being at the same time the smallest benchmark in this category.
On average, the number of reachable methods is reduced by \NUMBER{6.3\%}. 

For Renaissance, the biggest reduction is achieved on the \textit{chi-square} benchmark, which is reduced by \NUMBER{17.2\%}. The other Spark benchmarks \textit{als}, \textit{dec-tree}, and \textit{log-regression} are reduced by at least \NUMBER{15\%}.
The least impacted benchmark is \texttt{reactors}, reduced by \NUMBER{3.7\%}, again being also one of the smallest in this category.
In general, the whole suite is reduced by~\NUMBER{8.4\%}.

Overall, we can observe that bigger applications generally offer more potential for optimizations, because they have more libraries out of which only subsets are used and thus the remainder can be removed. 
Nevertheless, there are some significant outliers, for example, the \textit{DaCapo Sunflow} benchmark, suggesting that our analysis works especially well if specific code patterns are used, such as a~method call within a~provably unreachable branch or a~not needed default value for an optional argument presented in Section~\ref{sec:motivatingExample}.

\begin{figure*}
\centering

\begin{subfigure}{\textwidth}
    \includegraphics[width=\textwidth]{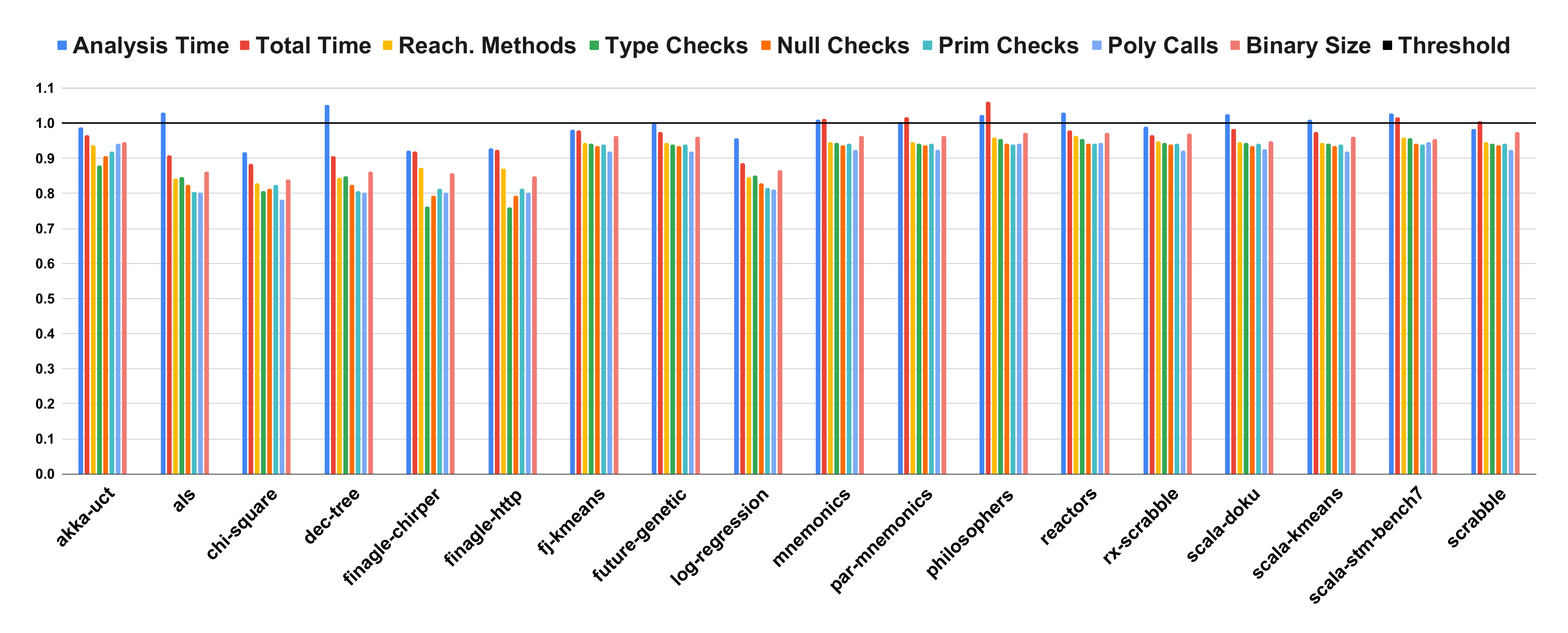}
    \caption{Renaissance}
\end{subfigure}
\hfill
\begin{subfigure}{0.49\textwidth}
    \includegraphics[width=\textwidth]{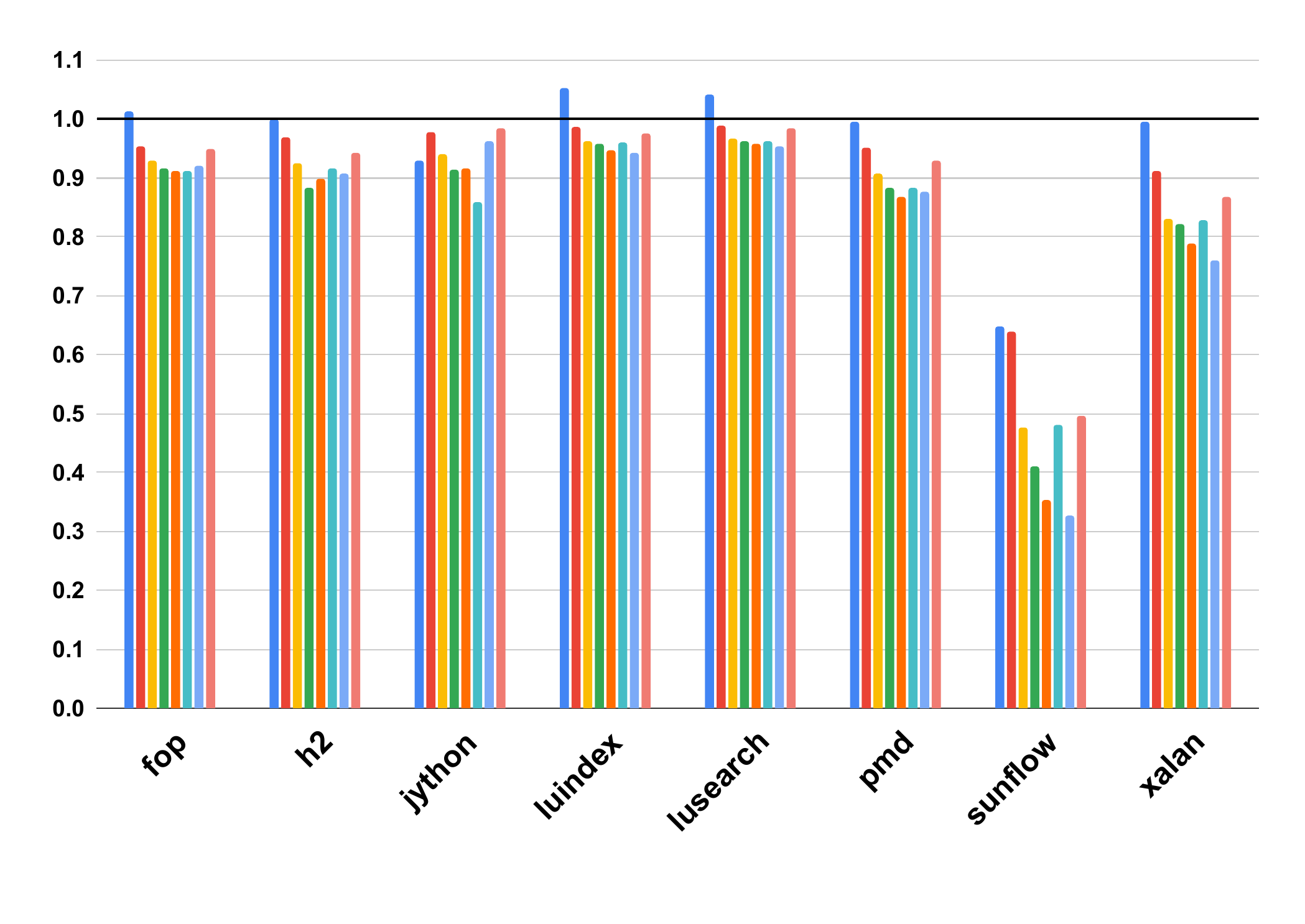}
    \caption{DaCapo}
\end{subfigure}
\hfill
\begin{subfigure}{0.49\textwidth}
    \includegraphics[width=\textwidth]{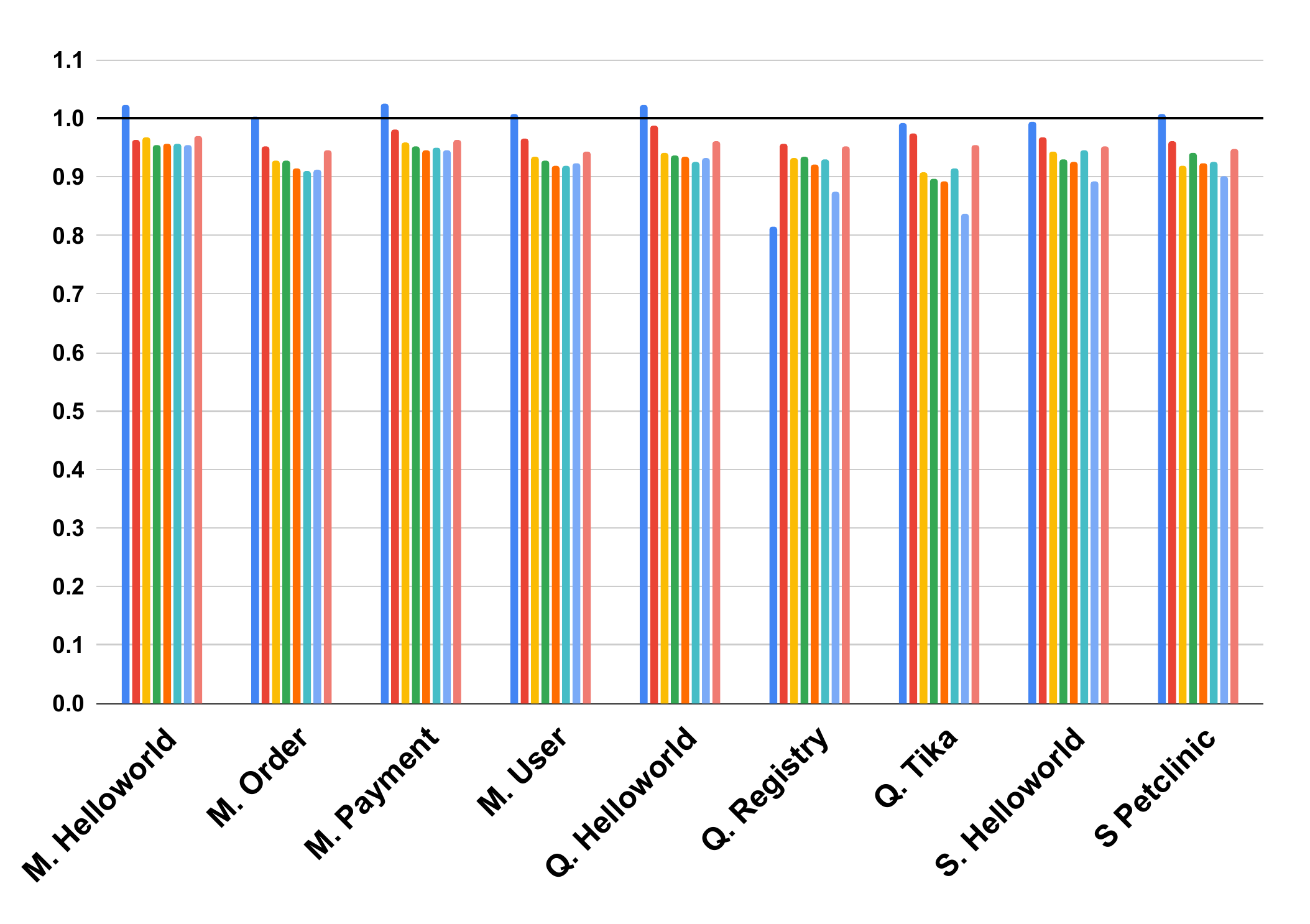}
    \caption{Microservices}
\end{subfigure}

\caption{Normalized metrics per bench suite. Lower is better, anything below 1.0 is an improvement. \analysisName{} consistently improves all the metrics apart from \textit{analysis time}, where the results are inconclusive. Nevertheless, even for \textit{analysis time} the average is still slightly better (\NUMBER{-1.6\%}) for \analysisName{}.}
\label{fig:normalized}
\end{figure*}


In Figure~\ref{fig:normalized}, we present all the metrics normalized to the baseline points-to analysis to provide a~quick overview complementing the detailed Table~\ref{tab:detailedtable}. 
Using the charts, we can quickly notice the already discussed \textit{DaCapo Sunflow} outlier and also the general trend of \NUMBER{9\%} reduction in \textit{reachable methods} across all the benchmarks.

\subsubsection*{Impact on Analysis Time}
Interestingly, we can also observe that the analysis time does not increase for most benchmarks. 
On the contrary, the average analysis time is even reduced by \NUMBER{1.6\%}, suggesting we can actually \textit{have a~free lunch} and use a~more precise analysis without negatively impacting the analysis time.
Indeed, even though our analysis is more complex compared to the baseline points-to analysis, it is also more precise, leading to fewer methods marked as reachable. 
Fewer reachable methods mean less work for the analysis, \dk{but also for the compilation that follows, which can be seen on the \textit{total time} metric, being reduced by \NUMBER{4.4\%} on average.}

There is a~tipping point after which the cost of a~more precise analysis outweighs the reduction due to fewer reachable methods. We believe our analysis is a~good spot in the design space, providing enough precision improvement to make a~difference without impacting the analysis time.

\dk{
\subsubsection*{Impact on Compiler Optimizations}
There are numerous optimizations that can benefit from the facts proved by \analysisName{}, including but not limited to: (1) \textit{Dead code elimination} -- flows that remain disabled until the end of the analysis correspond to instructions that can never be executed, so we can safely remove them. If a filtering flow has an empty value state after the analysis, the entire branch for which the given flow was generated is proved unreachable and can be removed; (2) \textit{Intraprocedural constant folding} -- if a method parameter is proved to be a constant value, it allows more intraprocedural constant folding after the analysis; (3) \textit{Method inlining} -- Due to the dead code elimination and constant folding, methods become smaller, making them more amendable to inlining, which can unlock further optimizations.

Our current evaluation focuses mainly on reducing the set of reachable methods. 
The \textit{counter metrics} show that \analysisName{} also reduces the number of type checks, null checks, primitive checks, and virtual invokes in the methods that remain reachable, but
more work is needed to evaluate the impact of \analysisName{} on runtime performance.
}

\dk{
\subsubsection*{Impact on Binary Size}
The reduction of reachable methods propagates to the binary size of the resulting applications. We can observe a~significant reduction in this metric as well, more precisely \NUMBER{max 5.6\%, min 3.1\%, avg 4.6\% for microservices, max 16.0\%, min 2.6\%, avg 7.3\% for Renaissance, max 50.5\%, min 1.6\% avg 11.0\% for DaCapo, and 7.4\% on average across all the benchmarks}.
In general, the binary size reduction follows a similar trend to the reduction in reachable methods.
}

\subsubsection*{Discussion}
Overall, we conclude that while \analysisName{} can also remove more type checks, null checks, primitive checks, and devirtualize more virtual calls, the biggest benefit compared to the baseline is \textit{reducing the total number of reachable methods without increasing the overhead in terms of analysis time}, even reducing the analysis time in many cases. 
Reducing the number of reachable methods not only speeds up the analysis itself, but it also reduces the workload of the compilation phase that follows the analysis\dk{, which can be observed on the \textit{total time} being \NUMBER{4.4\%} smaller on average}.
On top of that, fewer methods also mean less code to include in the compiled binary, thus reducing the binary size \dk{by \NUMBER{7.4\%} on average}. 
We believe these properties make our analysis a~suitable extension for a points-to analysis used in optimizing compilers.

\dk{The increased precision of \analysisName{} can also improve the results of any subsequent client interprocedural static analysis, such as a~taint analysis, data-flow analysis or abstract interpretation, as removing spurious edges from the call graph can lead to fewer false alarms.}

\section{Related Work}
\subsubsection*{Points-to Analysis}
Points-to analysis (PTA) has been applied in many areas including compilers~\cite{Wimmer19}, security analysis~\cite{Arzt14,Livshits05}, bug hunting~\cite{Guyer05,Liu18}, heap allocation analysis~\cite{Stancu15}, escape analysis~\cite{Lundberg09}, call graph construction~\cite{Karim12,Reif16}, and program understanding~\cite{Li16,Fiutem99}. 
Typically, PTA is implemented using inclusion-based constraint solving~\cite{andersen94,barbar21}, type system based unification~\cite{Steensgaard96}, or via reduction to a~graph reachability problem~\cite{Li11}. 

The frameworks implementing PTA essentially fall into three different categories:
1)~\emph{imperative}, e.g., Spark~\cite{Lhotak03}, WALA~\cite{Wala}, and Qilin~\cite{HeDongjie22} (written purely in Java), 
2)~\emph{declarative}, e.g., DOOP~\cite{Bravenboer09} (written in Datalog),
3)~\emph{hybrid}, e.g., Paddle~\cite{Lhotak08} (with a declarative core in Datalog and the rest of the infrastructure in Java).
\analysisName{} is implemented on top of an existing imperative points-to analysis framework, is parallel, utilizes inclusion based-solving, and its formal presentation is inspired by the works of He et al.~\cite{HeDongjie22}. 


\vspace{-2mm}
\subsubsection*{Context-sensitivity}
A~well-researched property of points-to analysis, and in general any interprocedural analysis, is context-sensitivity, which allows each method in the program to be analyzed under different contexts~\cite{Li18}.
Different types of context-sensitivity have been studied in the past~\cite{Kastrinis13,Milanova02,Smaragdakis11,sharir78,Thakur20}, e.g., call-site-sensitivity using method call-sites~\cite{sharir78} and object-sensitivity using allocation sites~\cite{Milanova02}.

While a~lot of effort has been made to improve the scalability of context-sensitive analysis~\cite{barbar21,LuJingbo19,Thakur20}, it still often scales poorly for large applications~\cite{Wimmer24}. 
As the analysis time is an~important metric for optimizing compilers, our implementation of \analysisName{} is context-insensitive.
However, the approach can be applied in a context-sensitive analysis too. 
\vspace{-2mm}
\subsubsection*{Partial Flow-sensitivity}
Roy et al.\ proposed partial flow-sensitivity~\cite{roy07}, which maintains flow-sensitivity only for a~specific set of program points. However, their approach expected the user to provide a~set of program points, requiring a~manual step before the analysis and thus rendering it unsuitable for optimizing compilers.

Wei et al.~\cite{wei14} presented a~partially flow-sensitive points-to analysis for Javascript using a~State Preserving Block Graph, a~transformed control-flow graph whose blocks are aggregated into region nodes according to whether or not they contain a~state-update statement, i.e., a property write or delete. 
Nevertheless, their approach still increased the overhead of the analysis compared to the baseline analysis.

The ideas of partial flow-sensitivity are arguably close to the principles of our analysis.
\analysisName{} could perhaps be seen as another representative of this category.
However, it is not only fully automated and capable of increasing precision but it also keeps the same overhead as a~baseline flow-insensitive analysis.
\vspace{-2mm}
\subsubsection*{Value Flow Analysis}
Value flow analysis has been successfully applied to finding many types of source-sink problems including null pointer derefences~\cite{Hovemeyer07}, memory leaks~\cite{Cherem07}, buffer overruns~\cite{Livshits03}, the usage of uninitialized variables~\cite{Ye18}, and concurrency issues~\cite{Cai21}. 
Sui et al.~\cite{Sui16} presented a~scalable and precise interprocedural static value flow analysis for C programs. 
Livshits et al. introduced IPSSA~\cite{Livshits03}, extending SSA with def-use relationships due to pointer dereferences and procedure calls.
Shi at al. presented Pinpoint~\cite{Shi18}, a~sparse value flow analysis using a~holistic approach in which the underlying points-to analysis is aware of the high-level properties being checked and computes only points-to sets that are relevant to checking these properties. 
However, such an on-demand approach cannot be directly utilized in optimizing compilers where the goal is to compute information about \textit{all program locations}.

Most value flow analysis implementations rely on SMT solving~\cite{Shi18,Shi21}. 
Formulas representing individual program paths are constructed and used to determine path feasibility~\cite{Cherem07}. 
In our approach, thanks to \textit{filtering flows} and \textit{predicate edges}, we can obtain better precision compared to the baseline analysis without constructing any formulas.
\vspace{-2mm}
\subsubsection*{Dataflow Analysis}
Dataflow analysis is a~framework that can be utilized for a~wide range of tasks, including points-to analysis or constant propagation~\cite{Sagiv96}. Often, dataflow analysis can also be reduced to a~graph problem~\cite{Reps95}.
Fischer et al.~\cite{Fischer05} presented predicates in dataflow analysis using \textit{predicated lattices} which partition the
program state according to a~set of predicates and track a~lattice element for each partition. 
The most distinguishing feature of \analysisName{} compared to classical dataflow analysis is the usage of \textit{predicate edges} instead of \textit{predicate latttices}, allowing us to keep using a~simple lattice. 
\vspace{-2mm}
\subsubsection*{Gated SSA form}
Our approach bears a~certain resemblance to the gated SSA form defined by Ottenstein et al.~\cite{ottenstein90}.
Contrary to the Gated SSA, we do not extend the $\phi$ nodes and instead use \textit{predicate edges} to capture the impact of branching conditions.
\vspace{-2mm}
\subsubsection*{Sparse Conditional Constant Propagation}
Sparse Conditional Constant Propagation (SCCP)~\cite{Wegman91,Click95} is a compiler optimization that propagates constants while taking into account the control flow of the program and the reachability of individual instructions. 
It has been traditionally applied only intraprocedurally on a single compilation unit (one root method and possibly many inlined callees), which leads to low precision especially when programs are built from many small functions, as is common in languages such as Java or C\#. 
While we presented our work as a~static analysis method, it can also be seen as a novel Whole Program Sparse Conditional Constant Propagation. 
While existing SCCP implementations also operate on a lattice of values, we are not aware of an implementation that uses a complex lattice for object values that would be comparable to a points-to-set. Within a single compilation unit, SCCP cannot handle complex operations on types. Instead, existing SCCP usually focuses on primitive values, using a lattice for primitive values that is more complicated than our simple constants-only representation of primitive values.

\section{Conclusions}
In this paper, we presented \analysisName{} that can be viewed as an extension of a~flow-insensitive points-to analysis that models primitive values and maintains flow-sensitivity for local variables and simple branching instructions only, while falling back to the flow-insensitive analysis for the rest, thus increasing the precision without significantly increasing the analysis overhead. 
We implemented our analysis in the GraalVM Native Image and evaluated it on a wide range of benchmarks including DaCapo, Renaissance, and a~set of microservice applications. 
Overall, our analysis reduces the number of reachable methods by \NUMBER{9\%} on average without negatively impacting the analysis time, in fact even reducing it in many cases. 
We believe our analysis is a~sweet spot in the design space, improving the precision of the analysis without negatively impacting analysis time.

In the future, one can think whether the amount of information considered by the analysis can be extended further without hurting the scalability.

\begin{acks}
We thank all members of the GraalVM team at Oracle Labs and the Institute for System Software at the Johannes Kepler University Linz for their support and contributions.
This work was supported by the Czech Science
Foundation project 23-06506S and the FIT BUT internal project FIT-S-23-8151.
Oracle and Java are registered trademarks of Oracle and/or its affiliates.
Other names may be trademarks of their respective owners.
\end{acks}

\appendix
\section{Artifact Appendix}
The Docker image that replicates our experiments~\cite{Artifact}, the results of which are presented in Section~\ref{section:Evaluation}, particularly in Table~\ref{tab:detailedtable} and Figure~\ref{fig:normalized}, is available on Zenodo for artifact evaluation. For more details, we refer to the artifact itself, which contains a~detailed README file.

\bibliographystyle{ACM-Reference-Format}
\IfFileExists{paper.bbl}{

}{\bibliography{literature}}

\section{Predicated Value Propagation Graph}
\label{sec:app:pvpg}
This section formalizes the \textit{predicated value propagation graph} (PVPG), the core data structure upon which \analysisName{} is executed. First, we define the base language that serves as the input to our analysis. Second, we define the lattice representing the values propagated through PVPGs. Third, we define the structure of PVPGs. Finally, we present an algorithm for creating a~PVPG by a~linear pass over a~program written in the base language. We show how a~PVPG is used in \analysisName{} in Section~\ref{sec:valuepropagationrules}.

\subsection{Base Language}
For the formal definition of our analysis, we use the base language presented in Figure~\ref{fig:baseLanguage} with instructions for object instantiation, field access, method invocation, and control flow using $\boldsymbol{\mathit{jump}}$ and $\boldsymbol{\mathit{if}}$. 
The instructions within a~method are separated into \textit{blocks}.

The base language is in SSA form, i.e., all variables have one static definition, all variables are defined before their first use, and all usages are dominated by the single definition.

Each method has exactly one $\boldsymbol{\mathit{start}}\:(p_0,...,p_n)$ instruction, at the beginning of the first block. 
This instruction explicitly defines method parameters $p_0,...,p_n$ where $p_0$ is the receiver object.
Without loss of generality, each method has only a single $\boldsymbol{\mathit{return}}\:v$ instruction. 

Blocks starting with a~$\boldsymbol{\mathit{merge}}\:[u \gets \phi_1(u_1,...,u_n),..., \:v \gets \phi_k(v_1,...,v_n)]\:m$ instruction represent control flow merges and can form loops.
A~$\boldsymbol{\mathit{merge}}$ instruction has a label $m \in M$, where $M$ is a set of labels.
A~$\boldsymbol{\mathit{merge}}$ instruction defines a~unique named location $m$ that is the target of $\boldsymbol{jump}$ instructions that use the same label. 
Furthermore, a~$\boldsymbol{\mathit{merge}}$ instruction defines a~set $\{\phi_1,...,\phi_k\}$ of $\phi$ instructions, one for each variable that has multiple reaching definitions at the control flow merge.
Every $\phi$ instruction has as many arguments as there are $\boldsymbol{\mathit{jump}}$ instructions to its $\boldsymbol{\mathit{merge}}$.

A~$\boldsymbol{\mathit{label}}$ $l$ instruction creates a~unique named location $l \in L$, where $L$ is a~second set of labels disjoint from $M$.
A~$\boldsymbol{\mathit{label}}$ instruction marks the beginning of one of the two branches of an~$\boldsymbol{\mathit{if}}$ instruction using the same label. 
A~block ending with an $\boldsymbol{\mathit{if}}$ instruction always has two successor blocks beginning with $\boldsymbol{\mathit{label}}$ instructions.
A~block beginning with a~$\boldsymbol{\mathit{label}}$ instruction always has a~single predecessor ending with $\boldsymbol{\mathit{if}}$, i.e., it can never be the target of a$~\boldsymbol{\mathit{jump}}$ instruction. 

The constraints above ensure that there are no \emph{critical edges}, i.e., control flow edges from a block with multiple successors to a block with multiple predecessors. 
This is without loss of generality as long as the language does not have computed jump targets. 

We explicitly support three types of conditions: $v_1 = v_2$, $v_1 < v_2$, and $v\:\boldsymbol{\mathit{instanceof}}\:T$. 
Furthermore, nullcheck is covered implicitly by a~$v_1 = v_2$ check where $v_2 \gets \mathtt{null}$. 
Boolean values are encoded as integers, and the truth test is then represented as a~$v_1 = v_2$ check where $v_2 \gets 1$.
Other conditions are excluded from the formal base language for simplicity.
Having only $<$ and $=$ as comparisons is without loss of generality because $>$, $\leq$ $\geq$, and $\neq$ can be expressed by switching the order of operands and/or the order of the $\boldsymbol{\mathit{then}}/\boldsymbol{\mathit{else}}$ blocks. 
The $n$ case in the \texttt{Expr} rule covers primitive literals.
Since we do not seek to model arithmetic computations, we use the generic $\boldsymbol{\mathit{Any}}$ instruction instead, representing any arithmetic.

\subsection{The Lattice Domain of \analysisName{}}
\label{sec:lattice}

\begin{figure}
\scriptsize
\centering
\resizebox{.48\textwidth}{!}{%
\begin{bnf}
$\:$ : \textsf{Method} ::= \textit{Block}*
;;

$\:$ : \textsf{Block} ::= \textit{BlockBegin} \textit{s}* \textit{BlockEnd}
;;

$\:$ : \textsf{BlockBegin} ::=
| $\boldsymbol{\mathit{start}}\:(p_0,...,p_n)$ 
| $\boldsymbol{\mathit{merge}}\:[u \gets \phi_1(u_1,...,u_n),..., \:v \gets \phi_k(v_1,...,v_n)]\:m$ 
| $\boldsymbol{\mathit{label}}\:l$ 
;;

$s$ : \textsf{Statement} ::=
| $v \gets e$ 
| $v \gets r.x$ 
| $r.x \gets v$ 
| $v \gets v_0.m(v_1,..,v_n)$ 
;;

$\:$ : \textsf{BlockEnd} ::=
| $\boldsymbol{\mathit{return}}\:v$ 
| $\boldsymbol{\mathit{jump}}\:m$ 
| $\boldsymbol{\mathit{if}}\:c\:\boldsymbol{\mathit{then}}\:l_{then}\:\boldsymbol{\mathit{else}}\:l_{else}$ 
;;

$c$ : \textsf{Cond} ::=
| $v_1 = v_2$  
| $v_1 < v_2$  
| $v\:\boldsymbol{\mathit{instanceof}}\:T$
;;

$e$ : \textsf{Expr} ::= 
| $n$
| $\boldsymbol{\mathit{Any}}$ 
| $\boldsymbol{\mathit{new}}\;T$ 
| \texttt{null}
\end{bnf}%
}
\vspace*{-3mm}
\caption{The base language considered by the analysis.}
\label{fig:baseLanguage}
\end{figure}

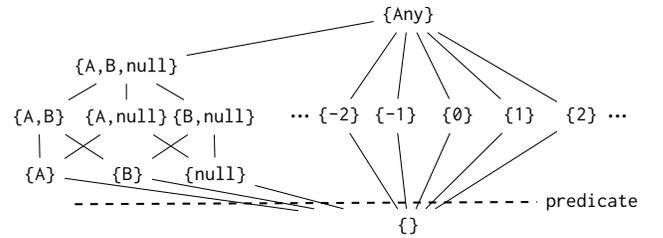
\begin{figure}
\centering
\resizebox{0.48\textwidth}{!}{%
\begin{tikzpicture}[sibling distance=1.2cm, level distance=0.70cm]

\node [xshift=0em] (topnode) at (0,5) { \footnotesize \texttt{\{\texttt{Any}\}}}
    child {
    node [xshift=-11em] { \footnotesize \texttt{\{A,B,null\}}} {
        child { 
            node [xshift=-0em] { \footnotesize \texttt{\{A,B\}}}
        }
        child { 
            node [xshift=-0em] { \footnotesize \texttt{\{A,null\}}}
        }
        child { 
            node [xshift=-0em] { \footnotesize \texttt{\{B,null\}}}
        }
    }
}
;

\node[minimum width=3cm, yshift=6em](bottomnode) { \footnotesize \texttt{\{\}} } [grow'=up]
child  {  
    node [xshift=1em] { \footnotesize \texttt{\{A\}} }
}
child  {  
    node [xshift=1em] { \footnotesize \texttt{\{B\}} }
}
child  {  
    node [xshift=1em]{ \footnotesize \texttt{\{null\}} }
}
child  {  
    node [xshift=0.9em, yshift=2.3em] { \Large ... } edge from parent[draw=none]
}
child  {  
    node [xshift=-1em, yshift=2.3em] { \footnotesize \texttt{\{-2\}} }
}
child  {  
    node [xshift=-2.2em, yshift=2.3em] { \footnotesize \texttt{\{-1\}} }
}
child  {  
    node [xshift=-3.2em, yshift=2.3em] { \footnotesize \texttt{\{0\}} }
}
child  {  
    node [xshift=-4.2em, yshift=2.3em] { \footnotesize \texttt{\{1\}} }
}
child  {  
    node [xshift=-5.2em, yshift=2.3em] { \footnotesize \texttt{\{2\}} }
}
child  {  
    node [xshift=-7.2em, yshift=2.3em] { \Large ... } edge from parent[draw=none]
}
;


\draw (topnode-1-1) -- (bottomnode-1);
\draw (topnode-1-1) -- (bottomnode-2);

\draw (topnode-1-2) -- (bottomnode-1);
\draw (topnode-1-2) -- (bottomnode-3);

\draw (topnode-1-3) -- (bottomnode-2);
\draw (topnode-1-3) -- (bottomnode-3);

\draw (topnode) -- (bottomnode-5);
\draw (topnode) -- (bottomnode-6);
\draw (topnode) -- (bottomnode-7);
\draw (topnode) -- (bottomnode-8);
\draw (topnode) -- (bottomnode-9);

\node (predLeft) [below of = bottomnode-1, xshift=1.0em, yshift=1.7em] {};
\node (predRight) [below of = bottomnode-10, yshift= -0.5em, xshift=-1em] {\footnotesize \texttt{predicate}};
\draw[dashed, thick] (predLeft) -- (predRight);

\end{tikzpicture}%
}
\vspace*{-3mm}
\caption{An example of the lattice $\mathbb{L}$ for $\mathbb{T}=\{\mathtt{A},\mathtt{B},\mathtt{null}\}$ and all the primitive values.
The line denoted as \texttt{predicate} indicates the fact that any non-empty value leads to the triggering of \textit{predicate} edges.}
\label{fig:latticeL}
\end{figure}

In this section, we define the lattice $\mathtt{L}_{\leq}$ whose values are propagated through PVPGs. 
Figure~\ref{fig:primitiveLattice} in Section~\ref{sec:primAbstractions} already introduced our primitive value lattice $\mathbb{P}$. 
To improve the scalability, we model neither sets nor intervals of values. 
The join of any two constant values results immediately in \texttt{Any}.
Following the work of Wimmer et al.~\cite{Wimmer24}, objects are represented by their types only and modelled using the subset lattice $\mathbb{S} = (2^{\mathbb{T}},\subseteq)$ over the set of program types $\mathbb{T}$.
The \textit{nullness} of a~given value is modelled using the special value $\mathtt{null} \in \mathbb{T}$. 

Formally, $\mathbb{L}_{\leq} = (\mathbb{L},\leq_{\mathbb{L}})$ where $\mathbb{L} = \{\{p\}\;|\;p \in \mathbb{P} \setminus \{\mathtt{Empty}\}\} \cup 2^{\mathbb{T}}$ and $\leq_{\mathbb{L}}$ is defined as the smallest relation such that (1) $\{p\} \leq_{\mathbb{L}} \{q\} \Leftrightarrow p \leq_{\mathbb{P}} q$ for any $p,q \in \mathbb{P} \setminus \{\mathtt{Empty}\}$,
(2) $S_1 \leq_{\mathbb{L}} S_2 \Leftrightarrow S_1 \leq_{\mathbb{S}} S_2$ for any $S_1,S_2 \in 2^{\mathbb{T}}$, 
(3) $\{\} \leq_{\mathbb{L}} S \leq_{\mathbb{L}} \{\mathtt{Any}\}$ for any $S \in 2^{\mathbb{T}}$, and 
(4) $\{\} \leq_{\mathbb{L}} \{p\} \leq_{\mathbb{L}} \{\mathtt{Any}\}$ for any $p \in \mathbb{P} \setminus \{\mathtt{Empty}\}$.
Note that $\top_{\mathbb{L}} = \{\top_{\mathbb{P}}\} = \{\mathtt{Any}\}$ and $\bot_{\mathbb{L}} = \bot_{\mathbb{S}} = \{\}$.
In $\mathbb{L}_{\leq}$, the primitive values are wrapped into 1-element sets (apart from \texttt{Empty} which is replaced by the empty set),
allowing us to treat all the \textit{value states} uniformly as sets in the text that follows.

The lattice $\mathbb{L}_{\leq}$ for $\mathbb{T} = \{\mathtt{A},\mathtt{B},\texttt{null}\}$ is given in Figure~\ref{fig:latticeL}.
The line denoted as \texttt{predicate} indicates the fact that any non-empty value leads to the triggering of \textit{predicate} edges.

In the rest of the text, since we do not have to distinguish between the lattice and its carrier set, we simply denote $\mathbb{L_{\leq}}$ as $\mathbb{L}$.

\subsection{Structure of a~PVPG}
A~PVPG models the flow of both primitive values and types interprocedurally. Vertices in a~PVPG are called \textbf{flows} to clearly distinguish them from the \textbf{instructions} in the base language.
In particular, the flows represent:
\begin{itemize}
    \item values of method parameters, variables, and fields read or written by a particular instruction (each instruction reading or writing to a variable produces a fresh flow),
    \item method calls, which also represent the returned value in the caller,
    \item values to be returned from a method call back to the caller,
    \item values of various conditions present in the code (including their negated and inverted versions as explained later),
    \item values resulting from joining the values of other flows (results of $\phi$  instructions),
    \item $\phi_{pred}$ flows for joining predicates,
    \item the always enabled predicate \predEnabled{}.
\end{itemize}
Each flow keeps a~reference to the underlying base language element it represents.
In the algorithms and inference rules below, we use the notation $\mathit{Flow}(i)$ to express the creation of a~new flow for a given base language element $i$, where $i$ can be either an instruction as a whole, a~condition, or a fresh~$\phi$ for the joining of values.
Subsequently, $f \colon Flow(i)$ denotes the flow $f$ created from $i$. 

The analysis computes the \textit{value state} of each flow, which represents a~conservative overapproximation of the values that can be assigned to the given flow during the runtime of the program. 
Flows can be connected via three types of \textbf{edges} described below.

A \textit{use} edge represents a~\textit{def-use} dependency between flows. 
If there is a~\textit{use} edge between flows \texttt{s} and \texttt{t}, denoted by \mbox{$\mathtt{s}\rightsquigarrow_{use}\mathtt{t}$}, the \textit{value state} of \texttt{t} has to be at least as big as the \textit{value state} of \texttt{s} in terms of the underlying lattice provided that $s$ is executable and no filtering is applied (the semantics of filtering is explained later).

A \textit{predicate} edge represents a~control-flow dependency. A~\textit{predicate} edge between flows \texttt{s} and \texttt{t}, denoted by \mbox{$\mathtt{s}\rightsquigarrow_{pred}\mathtt{t}$}, defines that if \texttt{s} is executable and has a~non-empty \textit{value state}, \texttt{t} is also executable. 
Every flow is the target of an incoming \textit{predicate} edge, apart from \predEnabled{}, which is always enabled and therefore does not need an incoming edge.
Flows that do not have any suitable predicate are assigned a~\textit{predicate} edge from \predEnabled{}.
A~$\phi_{pred}$ flow has multiple incoming \textit{predicate} edges. 
In that case, it suffices that any of its predicates has a~non-empty state to enable the execution of the target flow.
Apart from \predEnabled{} and $\phi_{pred}$, flows representing conditions and method calls can also be the sources of predicate edges.

An \textit{observe} edge represents an additional flow-specific dependency. An \textit{observe} edge between flows \texttt{s} and \texttt{t}, denoted as \mbox{$\mathtt{s}\rightsquigarrow_{obs}\mathtt{t}$}, defines that \texttt{t} has access to the \textit{value state} of \texttt{s}, and \texttt{t} is notified when the state of \texttt{s} changes to perform some flow-specific task such as a~field update or method resolution and linking (discussed more below).
Observe edges are created in three cases: (1) to link a~flow representing a~\textit{receiver} with subsequent flows representing instructions for calling methods, (2) to link a flow representing an object with subsequent flows representing instructions for loading/storing fields, (3) to connect the second argument of a filtering flow performing a~binary comparison, i.e. the argument according to whose \textit{value state} the filtering is performed (more on that later).

\subsection{Creating a~PVPG}
In this section, we describe how the PVPG for a~given method is created by a~sequential pass over the method body.
The rules for value propagation are defined later in Appendix~\ref{sec:valuepropagationrules}.
The PVPG is created by traversing the basic blocks in reverse postorder. 
The instructions within each basic block are processed sequentially from top to bottom.
During the traversal, flows are created for the encountered base language elements.
The traversal maintains a~state for each basic block consisting of:
(1) A~mapping \textit{m} from variables to previously created flows, which is used
to connect flows with their dependencies, e.g. to establish a \textit{use} edge between the flow representing \texttt{x} and a~subsequently created filtering flow \texttt{x != 0}, 
(2) a~reference denoted as \textit{pred} that is continuously updated to refer to the last encountered predicate.
The predicate referenced by \textit{pred} is used to establish predicate edges.
At the beginning of a~method, when no suitable flow is available yet, the special \predEnabled{} flow is used, which is always enabled, i.e. the initial flows inside methods are always enabled.

Initially, the mapping \textit{m} in each basic block is empty.
The $pred$ of each basic block is set depending on its initial instruction. 
For a~$\boldsymbol{\mathit{start}}$ instruction, \textit{pred} is set to \predEnabled{} .
For each basic block starting with a~$\boldsymbol{\mathit{merge}}$ instruction, \textit{pred} is set to a~fresh $\phi_{pred}$ flow, which represents the merge of predicates at the current program point (as discussed in Section \ref{sec:mergingValues}). 
Intuitively, a~basic block starting with a~$\boldsymbol{\mathit{merge}}$ is reachable iff the end of any of its predecessors is.
For a~$\boldsymbol{\mathit{label}}$ instruction, \textit{pred} is set when processing the predecessor basic block as it depends on the branching condition.

When processing an instruction, new flows and edges are created and the basic block variables \textit{m} and \textit{pred} are updated as described in Figure~\ref{tab:createPVPG}.
When processing a~$\boldsymbol{\mathit{start}}$ instruction, flows for method parameters are created.
$\boldsymbol{\mathit{merge}}$ and $\boldsymbol{\mathit{label}}$ instructions do not require any treatment as the state propagation between basic blocks is handled from the ends of predecessors.
Each ~$v \gets e$ instruction is handled by creating a~flow that represents $v$ and that is therefore stored in the mapping $b.m$.
Next, each $v \gets r.x$ instruction is handled similarly to $v \gets e$, with an additional \textit{observe} edge from $r$ because the load needs to be notified every time the \textit{value state} of the object $r$ changes. 
Each $r.x \gets v$ instruction has an additional \textit{use} edge from $v$ through which the values to be assigned to $r.x$ flow.
Each $v \gets v_0.m(v_1,..,v_n)$ instruction again needs an \textit{observe} edge for the receiver $v_0$.
Note that the call also represents the returned result $v$, which may be empty if the method never returns. 
To handle this case precisely, $v \gets v_0.m(v_1,..,v_n)$ becomes the new predicate.

\begin{figure*}
\footnotesize	
\begin{tabular}{lllll}
\toprule
Processed Instruction $i$ & Created Flows & Use Edges & Observe Edges & Modifications of \textit{m} and \textit{pred} \\
\midrule
$\boldsymbol{\mathit{start}}\:(p_0,...,p_n)$ & $\forall j \in [0,n]: f_j \gets \mathit{Flow}(p_j)$ &  &  &  $\forall j \in [0,n]: \mathit{b.m}[p_j] \gets f_j$ \\
$\boldsymbol{\mathit{merge}},\boldsymbol{\mathit{label}}$ & & & & \\
$v \gets e$ & $f \gets \mathit{Flow}(i)$ & &  & $\mathit{b.m}[v] \gets f$ \\
$v \gets r.x$ &  $f \gets \mathit{Flow}(i)$ &  & $\mathit{b.m}[r]\rightsquigarrow_{obs}f$ & $\mathit{b.m}[v] \gets f$ \\
$r.x \gets v$ & $f \gets \mathit{Flow}(i)$ & $\mathit{b.m}[v]\rightsquigarrow_{use}f$ & $\mathit{b.m}[r]\rightsquigarrow_{obs}f$ & \\
$v \gets v_0.m(v_1,..,v_n)$ & $f \gets \mathit{Flow}(i)$ & & $\mathit{b.m}[v_0]\rightsquigarrow_{obs}f$ & $\mathit{b.m}[v] \gets f, \mathit{b.pred} \gets f$ \\
$\boldsymbol{\mathit{return}}\:v$ & $f \gets \mathit{Flow}(i)$ & $\mathit{b.m}[v]\rightsquigarrow_{use}f$ & & \\
\midrule
$\boldsymbol{\mathit{jump}}\:m$ & \multicolumn{4}{l}{$\mathit{propagate}(b,\mathit{m.target})$} \\
\midrule
$\boldsymbol{\mathit{if}}\:c\:\boldsymbol{\mathit{then}}\:l_{then}\:\boldsymbol{\mathit{else}}\:l_{else}$ & $\mathit{initBlock}(b,l_{then},c),\mathit{initBlock}(b,l_{else},\mathit{inv}(c))$ \\
\bottomrule
\end{tabular}
\caption{The effect of processing each instruction \textit{i} when creating a~PVPG. The current basic block is denoted as $b$. Additionally, each flow $f$ is assigned a~predicate edge $\mathit{b.pred}\rightsquigarrow_{pred}f$ upon its creation. Note that \textit{Flow} is not a function in the mathematical sense but rather a~constructor creating a~new flow upon every invocation. The notation $\forall j \in [0,n]$ used in the first line denotes that a~set of flows indexed by integers 0 to n is handled uniformly. }
\label{tab:createPVPG}
\end{figure*}

$\boldsymbol{\mathit{jump}}$ instructions need to propagate the state of the current basic block $b$ to the successor $t$, which is done in the auxiliary function \textit{propagate} presented in Figure \ref{alg:propagate}.
First, a~\textit{predicate} edge is established between the current predicate in $b$ and the predicate of $t$, encoding the fact that if the end of $b$ is reachable, so is the beginning of $t$.
Second, the content of $b.m$ is propagated into $t.m$. 
Each variable $v \in b.m$ is handled as follows (note that we use the notation $\boldsymbol{\mathit{for}}\:v \in \mathit{b.m}$ to denote an iteration over all the variables, i.e. keys, in the mapping $m$). 
If $t$ does not yet have a~mapping for $v$, it is inherited from $b$.
Otherwise, we check if there is a collision, i.e. $b$ and $t$ have different flows representing $v$.
If not, nothing needs to be done.
If there is a collision, the PVPG has to be extended as follows.
If the mapping for $v$ in $t$ is not a representation of a~$\phi$ instruction but just a~value inherited from some other predecessor,
we create a new $\phi$ flow $f$ and adjust the edges so that $f$ represents the joining of the values from the predecessors. 
Otherwise, the $\phi$ flow has already been created before, and we can simply add a~new \textit{use} edge to it.

\begin{figure}
\centering
\footnotesize
\begin{algorithmic}
\Function{\textit{propagate}}{b,t}
\State $\mathit{b.pred}\rightsquigarrow_{pred}\mathit{t.pred}$
\For{$v \in \mathit{b.m}$} \Comment{Iterate over all vars, i.e. keys, in $b.m$}
    \If{$v \notin \mathit{t.m}$} \Comment{Is there a mapping for $v$ in $t.m$?}
        \State $\mathit{t.m}[v] \gets \mathit{b.m}[v]$
    \Else
        \State $b_v \gets \mathit{b.m}[v]$
        \State $t_v \gets \mathit{t.m}[v]$
        \If{$b_v \neq t_v$} \Comment{Is there a collision (two flows for the same $v$)?}
            \If{\textbf{not} $\mathit{isPhi(t_v)}$} \Comment{No $\phi$ flow created yet?}
                \State $\mathit{t.m}[v] \gets \mathit{Flow(\phi)}$ \Comment{Create a fresh $\phi$ flow}
                \State $\mathit{t.pred}\rightsquigarrow_{pred}\mathit{t.m}[v]$
                \State $b_v\rightsquigarrow_{use}\mathit{t.m}[v]$
                \State $t_v\rightsquigarrow_{use}\mathit{t.m}[v]$
            \Else
                \State $b_v\rightsquigarrow_{use}t_v$ \Comment{$\phi$ flow already created, just add a~\textit{use} edge}
            \EndIf
        \EndIf
\EndIf
\EndFor
\EndFunction
\end{algorithmic}
\caption{The \textit{propagate} function handling $\boldsymbol{\mathit{jump}}$ instructions.}
\label{alg:propagate}
\end{figure}

For \textbf{if} instructions, it is necessary to initialize the state of both of their successor basic blocks.
To do so, the helper function \textit{initBlock} is called, which dispatches to the \textit{initUnary} and \textit{initBinary} functions presented in Figure~\ref{alg:initBlock} based on the type of the condition $c$.
Notice that the condition is inverted when calling \textit{initBlock} for the \texttt{else} branch.
For a~unary condition (a~type check), a~new flow is created and inserted into the PVPG to represent the value of the checked variable $x$ after filtering it based on $c$.
For a~binary condition, both the left variable denoted as $c.l$ and the right variable denoted as $c.r$ involved in the condition are filtered, with the resulting flows stored into $t.m[c.l]$ and $t.m[c.r]$, respectively.  
The filtering of $c.l$ is similar to the handling of a unary condition, with an additional \textit{observe} edge for $c.r$ because the filtering flow needs to be notified in case any of its operands change.
The filtering of $c.r$ is a mirrored version of the process for $c.l$.
Note that the condition is \textit{flipped}, not inverted, i.e. $\mathit{flip}(<)=\;>$, while $\mathit{inv}(<)=\;\geq$.
This is because when filtering $y$ with respect to $x < y$, the filtering should only allow values of $y$ strictly greater than $x$.
The predicates are chained so that $\mathit{b.pred}\rightsquigarrow_{pred}f_l\rightsquigarrow_{pred}{f_r}$, where $f_l$ and $f_r$ are the flows created for the original condition and the flipped version, respectively. Furthermore, $f_r$ becomes the new predicate in $t$, ensuring that $t$ is considered reachable only iff the end of $b$ is reachable and the results of filtering of both $x$ and $y$ are not empty.

\begin{figure}
\centering
\footnotesize
\begin{subfigure}[t]{0.20\textwidth}
\begin{algorithmic}
\Function{\textit{initBlock}}{b,t,c}
    \If{$\mathit{isUnary(c)}$}
        \State $\mathit{initUnary}(b,t,c)$
    \Else
        \State $\mathit{initBinary}(b,t,c)$
    \EndIf
\EndFunction
\Function{\textit{initUnary}}{b,t,c}
    \State $f \gets \mathit{Flow}(c)$
    \State $\mathit{b.pred}\rightsquigarrow_{pred} f$
    \State $\mathit{b.m[c.x]}\rightsquigarrow_{use} f$
    \State $\mathit{t.m}[c.x] \gets f$
    \State $\mathit{t.pred} \gets f$
\EndFunction
\end{algorithmic}
\end{subfigure}
\begin{subfigure}[t]{0.20\textwidth}
\begin{algorithmic}
\Function{\textit{initBinary}}{b,t,c}
    \State $f_l \gets \mathit{Flow}(c)$
    \State $\mathit{b.pred}\rightsquigarrow_{pred} f_l$
    \State $\mathit{b.m[c.l]}\rightsquigarrow_{use} f_l$
    \State $\mathit{b.m[c.r]}\rightsquigarrow_{obs} f_l$
    \State $\mathit{t.m}[c.l] \gets f_l$
    \State $f_r \gets \mathit{Flow}(\mathit{flip}(c))$
    \State $f_l\rightsquigarrow_{pred} f_r$
    \State $\mathit{b.m[c.r]}\rightsquigarrow_{use} f_r$
    \State $\mathit{b.m[c.l]}\rightsquigarrow_{obs} f_r$
    \State $\mathit{t.m}[c.r] \gets f_r$
    \State $\mathit{t.pred} \gets f_r$
\EndFunction
\end{algorithmic}
\end{subfigure}
\caption{The \textit{initBlock}, \textit{initUnary}, and \textit{initBinary} functions handling $\boldsymbol{\mathit{if}}$ instructions. In the \textit{initBinary} function, the notation $c.l$ and $c.r$ is used to denote access to the left and the right argument of the comparison $c$, respectively.}
\label{alg:initBlock}
\end{figure}

\section{Value Propagation through PVPGs}
\label{sec:valuepropagationrules}
This section presents the core analysis algorithm based on the PVPG data structure. 
Given a~PVPG, primitives and types are propagated from source flows along the \textit{use} edges.
Each flow has a \textit{value state} describing the set of values that can be assigned at runtime to the code element, e.g., a~variable or a~field, represented by the given flow.
Contrary to a typical pointer assignment analysis, values in PVPG are only propagated by flows that are \textit{enabled} by their predicate. 
The values propagated through a~PVPG are from the lattice $\mathbb{L}$ described in Appendix~\ref{sec:lattice}.

We now proceed by defining \analysisName{} through a~series of inference rules precisely specifying the conditions for value propagation. 
Let $\mathbb{T}$, $\mathbb{M}$, $\mathbb{F}$, and $\mathbb{N}$ be pairwise disjoint sets representing the domains of types (including \texttt{null}), methods, field names, and PVPG flows, respectively.
Two auxiliary functions are used: $\mathit{LookUp} \colon \mathbb{T} \times \mathbb{F} \rightharpoonup \mathbb{N} $, which returns the flow representing the given \texttt{field} of the given \texttt{type}, and $\mathit{Resolve} \colon \mathbb{T} \times \mathbb{M} \rightharpoonup \mathbb{M}$, which resolves a virtual method invocation for the given \texttt{type} and \texttt{method}\footnote{Both \textit{LookUp} and \textit{Resolve} are partial, i.e. they only return values for valid combinations of input parameters.
We assume that the base language is well-typed. 
Invalid code trying to access a non-existent field or calling a non-existent method would be rejected by a~type system prior to running \analysisName{}.}.

The analysis computes the set of reachable methods $\mathbb{R} \subseteq \mathbb{M}$, the set of enabled flows $\mathit{Enabled} \subseteq \mathbb{N}$, the input \textit{value state} of each flow $\mathit{VS_{in}} \colon \mathbb{N} \rightarrow \mathbb{L}$, and the output \textit{value state} of each flow $\mathit{VS_{out}} \colon \mathbb{N} \rightarrow \mathbb{L}$.

Note that we introduce a separation between input and output \textit{value states}.
However, this concept is introduced purely to simplify the definition of some of the inference rules.
The actual implementation uses one \textit{value state} per flow.
All the sets above as well as all the \textit{value states} are initially empty.
The analysis is started by inserting a~set of \textit{root methods} into $\mathbb{R}$ and \predEnabled{} into $\mathit{Enabled}$. 
The set of \textit{root methods} contains the entry points from which the analysis is started, e.g., the \texttt{main} method.
Recall that \predEnabled{} is used as the special always-enabled predicate that is assigned to the flows at the beginning of the first block in a~method when no other predicate is available yet.
The analysis adds transitively reachable methods to $\mathbb{R}$ until a~fixed point is reached. 
The actual logic of the analysis is defined by the inference rules presented in Figure \ref{fig:inferenreFules}. 

\begin{figure*}
\scriptsize
\[
\inference{
    f \colon Flow(v \gets e),\; f \in \mathit{Enabled}
} {
    \mathit{VS_{in}}(f) = \begin{cases}
 n & \quad \text{for}\:v \gets n \\
 \texttt{Any} & \quad \text{for}\:v \gets \boldsymbol{\mathit{Any}} \\
 T & \quad \text{for}\:v \gets \boldsymbol{\mathit{new}}\;T \\
 null & \quad \text{for}\:v \gets \texttt{null} \\
\end{cases}
}{\text{\texttt{[Source]}}}
\quad
\inference{
    f_s \in \mathit{Enabled},\; f_s \rightsquigarrow_{use} f_t
} {
    \mathit{VS_{out}(f_s)} \leq_{\mathbb{L}} \mathit{VS_{in}(f_t)}
}{\text{\texttt{[Propagate]}}}
\quad
\inference{
    f_s \in \mathit{Enabled},\; \mathit{VS_{out}(f_s)} \neq \{\},\\
    f_s \rightsquigarrow_{pred} f_t
} {
    f_t \in \mathit{Enabled}
}{\text{\texttt{[Predicate]}}}
\]
\[
\inference{
    f \colon Flow(v \gets r.x),\; f \in \mathit{Enabled} \\
    t \in \mathit{VS_{out}(r)},\;\mathit{f_{t.x}} = \mathit{LookUp(t,x)}
} {
    \mathit{VS_{out}(f_{t.x})} \leq_{\mathbb{L}} \mathit{VS_{in}(f)}
}{\text{\texttt{[Load]}}}
\quad
\inference{
    f \colon Flow(r.x \gets v),\; f \in \mathit{Enabled} \\ 
    t \in \mathit{VS_{out}(r)},\; \mathit{f_{t.x}} = \mathit{LookUp(t,x)} \\
} {
    \mathit{VS_{out}(f)} \leq_{\mathbb{L}} \mathit{VS_{in}(f_{t.x})}
}{\text{\texttt{[Store]}}}
\quad
\inference{
    f \colon Flow(v \gets v_0.m(v_1,..,v_n)) \\
    f \in \mathit{Enabled},\;\; t \in \mathit{VS_{out}}(\mathit{v}_0) \\
    r = \mathit{Resolve(t,m)}
} {
\begin{gathered}
    \vspace{-4pt}
    r \;\mathit{added\;to}\; \mathbb{R},\; \forall i \in [0,n]\colon \mathit{v_i} \rightsquigarrow_{use} \mathit{p_i^{r}}\\
    \vspace{-0pt}
    \mathit{ret_{r}} \rightsquigarrow_{use} \mathit{f}
\end{gathered}
}{\text{\texttt{[Invoke]}}}
\]
\vspace{1em}
\[
\inference{
    f \colon Flow(c),\; f \in \mathit{Enabled},\; \mathit{isTypeCheck(f)} \\
    v_{out} = \{t\:|\:t \in \mathit{VS_{in}}(f) \land \mathit{c(t)}\}
}{
    v_{out} \leq_{\mathbb{L}} \mathit{VS_{out}(f)}
}{\text{\texttt{[TypeCheck]}}}
\quad
\inference{
    f \colon Flow(c),\; f \in \mathit{Enabled},\; \mathit{isComparison(f)} \\
    v_{out} = \mathit{Compare(c,VS_{out}(c.l),VS_{out}(c.r))}
}{
    v_{out} \leq_{\mathbb{L}} \mathit{VS_{out}(f)}
}{\text{\texttt{[Cond]}}}
\quad
\inference{
    \neg(\mathit{isTypeCheck(f)}), \\
    \neg(\mathit{isComparison(f)}), \\
    f \in \mathit{Enabled}
}{
    \mathit{VS_{in}(f)} \leq_{\mathbb{L}} \mathit{VS_{out}(f)}
}{\text{\texttt{[PassThrough]}}}
\]
\vspace{1em}
\[
\mathit{Compare(cond,v_l,v_r)} = 
    \begin{cases}
     \{\} & \quad \text{if } v_l = \{\} \lor v_r = \{\} \\
     \mathit{min_\mathbb{L}(v_l,v_r)}  & \quad \text{else if } \mathit{cond}\;\boldsymbol{is}\;\text{'='} \land\;(\mathtt{Any} \in v_l \lor \mathtt{Any} \in v_r) \\
     v_l \cap v_r & \quad \text{else if } \mathit{cond}\;\boldsymbol{is}\;\text{'='} \\
     v_l \setminus v_r & \quad \text{else if } \mathit{cond}\;\boldsymbol{is}\;\text{'$\neq$'} \\
     v_l & \quad \text{else if } \mathtt{Any} \in v_l \lor \mathtt{Any} \in v_r \\
     \{l\:|\:l \in v_l \land r \in v_r \land cond(l,r) \}        & \quad \text{otherwise}
    \end{cases}
\]
\caption{Inference rules used by \analysisName{}. In the rules, the meaning of the notation $f \colon Flow(v \gets e)$ denotes that the flow $f$ \emph{was} created from the base language instruction $v \gets e$ and similarly for the other instructions. Likewise, we use the notation $f \colon \mathit{Flow}(c)$ to denote a fact that a flow \emph{was} created from a condition $c$. The helper functions \textit{isTypeCheck} and \textit{isComparison} return \texttt{true} iff the flow was created from a~type check or a~comparison operator, respectively. }
\label{fig:inferenreFules}
\end{figure*}

The \texttt{Source} rule specifies the behaviour of a flow representing a~$v \gets e$ instruction. Once enabled, the \textit{value state} of $f$ contains the result of the evaluation of the expression $e$.

The \texttt{Propagate} rule specifies the conditions for propagating values through a~PVPG.
Observe that $\leq_{\mathbb{L}}$ is established between the \textit{value states} of $f_s$ and $f_t$.
This kind of formula is used throughout the rules to model any type of value propagation: Once the analysis reaches a fixed point, the \textit{value state} of $f_t$ has to be at least as big as the \textit{value state} of $f_s$ in terms of the lattice $\mathbb{L}$.

The \texttt{Predicate} rule defines predicate handling.
A~flow is enabled when its predicate is enabled and the predicate's \textit{value state} is non-empty.

The \texttt{Load} and \texttt{Store} rules represent field manipulation. 
Both rules operate on the flow $r$ representing the accessed object and lookup the flows representing fields named $x$ on all types $t$ in the \textit{value state} of $r$, denoted as $f_{t.x}$.
Recall that $Flow(v \gets r.x)$ represents the joined value of all the given fields $x$ for all the types in the \textit{value state} of the \texttt{receiver}. 
Therefore, the \texttt{Load} rule establishes $\leq_{\mathbb{L}}$ between the $\mathit{VS_{out}}$ of all the related $f_{t.x}$ and the $\mathit{VS_{in}}$ of the $Flow(v \gets r.x)$ itself.
Similarly, the \texttt{Store} rule establishes $\leq_{\mathbb{L}}$ between the $\mathit{VS_{out}}$ of $Flow(r.x \gets v)$ and the $\mathit{VS_{in}}$ of all the related $f_{t.x}$. 

The rules \texttt{TypeCheck}, \texttt{Cond}, and \texttt{PassThrough} define how the $\mathit{VS_{in}}$ of each flow is mapped to $\mathit{VS_{out}}$.
If the given flow represents a type check $c$, which can be either an $\boldsymbol{\mathit{instanceof}}$ check or its negated version, the $\mathit{VS_{out}}$ contains only the types from $\mathit{VS_{in}}$ that pass the condition $\mathit{c}$. 
If the condition $c$ is a~comparison, the auxiliary function \textit{Compare} is used to compute the $\mathit{VS_{out}}$.
If a~flow does not have any condition, its input is passed directly to its output without any filtering.

The auxiliary \textit{Compare} function uses an operator \textit{cond}, and $v_l$, $v_r$ as operands, returning the content of $v_l$ filtered with respect to $c$ and $v_r$.
If at least one of the operands is empty, we return an empty set because both operands are needed to perform the filtering.
If the operator is '$=$' and at least one of the operands contains \texttt{Any}, then the result of the filtering is the lower value in terms of the Lattice $\mathbb{L}$, e.g.:
\begin{gather*}
\mathit{Compare(\text{'='},\{\mathtt{Any}\},\{5\})} = \{5\}, \\
\mathit{Compare(\text{'='},\{\mathtt{Any}\},\{\mathtt{Any}\})} = \{\mathtt{Any}\}.
\end{gather*}
If the operator is '$=$', but none of the operands contains \texttt{Any}, the result is the set intersection of its arguments, e.g.:
\begin{gather*}
\mathit{Compare(\text{'='},\{\mathtt{A}, \mathtt{B}\},\{\mathtt{B}, \mathtt{C}}\}) = \{\mathtt{B}\}, \\ 
\mathit{Compare(\text{'='},\{3\},\{3\}) = \{3\}}, \\
\mathit{Compare(\text{'='},\{3\},\{5\}) = \{\}}.
\end{gather*}
If the operator is '$\neq$', the result is a~set difference: 
\[\mathit{Compare(\text{'$\not=$'},\{\mathtt{A}, \mathtt{B}\},\{\mathtt{B}, \mathtt{C}}\}) = \{\mathtt{A}\}.\]
This again works on primitives as well:
\begin{gather*}
\mathit{Compare(\text{'$\neq$'},\{\mathtt{0}\},\{\mathtt{0}\})} = \{\mathtt{}\}, \\ 
\mathit{Compare(\text{'$\neq$'},\{5\},\{\mathtt{3}\})} = \{5\}.
\end{gather*}
\balance
The remaining cases are '$<$', '$\leq$', and other relational operators apart from '$=$' and '$\neq$' handled above.
These operators are defined on primitives only, and so we know that $v_l$ and $v_r$ are the 1-element sets containing some primitive value, or \texttt{Any}.
If one of the operands contains \texttt{Any}, we cannot do any useful filtering, regardless of whether \texttt{Any} is in $v_l$ or $v_r$.
If $\mathtt{Any} \in v_l$, we could only reduce $v_l$ to an interval (assuming $\mathtt{Any} \notin v_r$), which we have decided not to use for scalability purposes.
If $\mathtt{Any} \in v_r$, we do not have enough information to filter $v_l$. 
Therefore, we simply return $v_l$ in both cases.
Finally, if none of the operands contains \texttt{Any}, we return the value from $v_l$ if it passes the \textit{cond} with respect to $v_r$; otherwise, we return the empty set:
\begin{gather*}
\mathit{Compare(\text{'<'},\{3\},\{5\}) = \{3\}}, \\ 
\mathit{Compare(\text{'<'},\{3\},\{1\}) = \{\}}.
\end{gather*}
Finally, the \texttt{Invoke} rule specifies how methods are linked. 
Every time a~new type is added to the \textit{value state} of the receiver $v_0$, it is passed to the \texttt{Resolve} function to determine the callee $r$.
Each callee $r$ is added to the set of reachable methods $\mathbb{R}$ and subsequently linked by creating a~\textit{use} edge from the argument flows $\mathit{v_i}$ in the caller to the parameter flows $\mathit{p_i^{r}}$ of the callee, and from the return flow $\mathit{ret_r}$ of the callee to the invoke $f$ itself, which represents the returned value in the caller.

\end{document}